\documentclass[onecolumn,pdflatex,nofootinbib,superscriptaddress,floatfix]{revtex4-1}
\usepackage[T1,T2A]{fontenc}
\usepackage[utf8]{inputenc}
\usepackage{graphicx}
\usepackage{bm}
\usepackage{placeins}
\usepackage{xspace}
\usepackage{bm,amsmath,amssymb,amsfonts,gensymb}
\usepackage{comment}
\usepackage{hyperref}
\hypersetup{
    colorlinks=true,
    linkcolor=blue,
    citecolor=blue,
    filecolor=blue,      
    urlcolor=blue,
}
\usepackage{cleveref}

\usepackage[russian,english]{babel}

\newcommand{\be}{\begin{equation}}
\newcommand{\ee}{\end{equation}}

\usepackage{color}

\begin{document}


\title{
 Thin accretion discs around spherically symmetric configurations with nonlinear scalar fields
 }
\author{O.~S.~Stashko}
\affiliation{Taras Shevchenko National University of Kyiv, Ukraine}
\author{V.~I.~Zhdanov}
\affiliation{Taras Shevchenko National University of Kyiv, Ukraine}
\author{A.~N.~Alexandrov}
\affiliation{Taras Shevchenko National University of Kyiv, Ukraine}

\date{\today}

\pacs{11}


\begin{abstract}

We study stable circular orbits (SCO) around   static spherically symmetric configuration of General Relativity with a non-linear scalar field (SF). The configurations are described by solutions of the Einstein-SF equations with  monomial SF potential  $V(\phi)=|\phi|^{2n}$, $n>2$, under the conditions of the asymptotic flatness and behavior of SF $\phi\sim 1/r$ at spatial infinity. We proved that under these conditions the solution exists and  is uniquely defined by the configuration mass $M>0$ and scalar "charge" $Q$. The solutions and the space-time geodesics have been  investigated numerically in the range $n\le40$, $|Q|\le 60$, $M\le60$. We  focus on how nonlinearity of the field affects  properties of  SCO distributions (SCOD), which in  turn  affect topological form of the thin accretion disk around the configuration. Maps are presented showing the location of possible SCOD types for different $M,Q,n$. We found many   differences from the Fisher-Janis-Newman-Winicour metric (FJNW) dealing with the linear SF, though basic qualitative properties of the configurations have much in common with the FJNW case. For some values of $n$, a topologically new SCOD type was discovered that is not available for the FJNW metric. All images of  accretion disks have a dark spot in the center (mimicking an ordinary black hole), either because there is no SCO near the center  or because of the strong deflection of photon trajectories near the singularity.

\end{abstract}

\maketitle

\section{Introduction}
 Scalar field (SF) configurations in  General Relativity   and its modifications  are interesting for several reasons. Various  SF models are extensively used in cosmology \cite{2006Copeland,Bamba_2012,novosyadlyj2015dark,linde2014inflationary}. Some  propositions to relax the well-known "Hubble tension"  involve SF in diverse  approaches to dynamical dark energy (see \cite{divalentino2021realm} for a review).  It is currently unknown,  whether the cosmological fields of the late epoch (if any) are the same as the fields that caused the early inflation, or they are of a completely different nature. In any case, the question arises about possible  manifestations of SF in relativistic astrophysical objects. Interest in alternative   models of these objects has  increased  significantly after the image of the accretion disk in the core of  M87 had been  obtained with the Event Horizon Telescope (EHT) \cite{Akiyama2019}, demonstrating  future  prospects to distinguish the black holes from their exotic mimickers. 

The first example of the static spherically symmetric solution of  Einstein equations with linear massless SF has been found by Fisher \cite{Fisher} and later, in the other form, by Janis, Newman \& Winicour \cite{JNW}  (hereinafter FJNW solution; see also \cite{PhysRevD.24.839,Virbhadra_1997}). The FJNW metric does not describe a black hole (BH), it has naked singularity (NS) in the center not hidded by an event horison. This is a typical feature of static solutions  with SF describing compact objects due to the  Bekenstein theorems \cite{Bekenstein1972, Bekenstein1972a}, see also a generalisation in case of multiple SFs \cite{Doneva_2020}.  The occurrence of NS in the real Universe is forbidden by the Penrose  Cosmic Censorship hypothesis \cite{Penrose1965,Penrose2002}; though,  the question on the validity of this hypothesis remains open \cite{Christodoulou1984,Ori1987,Joshi1993}. Current discussions on this topic have shifted to  issues of stability and realistic choices of initial data for the gravitational collapse that may or may not lead to NS (see, e.g.,  \cite{Ong2020} for a review). 

Anyway, the final answer regarding the role of SF  in astrophysics   must be based on observations. It should be noted that there are "exotic" structures \cite{Olivares_2020,vanaelst2021orbits,Herdeiro_2021,Abdikamalov_2019,Banerjee_2020,Banerjee_2020PhRvD,Sau_2020,Vincent_2021}, which mimic the  BHs yielding image similar to that observed by EHT \cite{Akiyama2019}. New theoretical efforts as well as observations with better resolution are mandatory.
Given the progress in astronomical technology, it is important to study in detail the properties of relativistic astrophysical objects, which can help to select the appropriate options from a variety of theoretical models. 
The main source of observational information from these objects is associated with the  distribution of surrounding radiating matter  (accretion disks, jets etc) and images of this matter seen by a distant observer. The very first step is to study  stable circular orbits (SCO) of test bodies and their distribution in gravitational field of the configuration. There are a number of papers  on this subject including those, which use FJNW solution dealing with the linear SF  \cite{Bambhaniya_2019,Sau_2020,Shaikh_2019,Gyulchev_2019,Gyulchev_2020,Chowdhury_2012,Zhou_2015} and it would be interesting to study effects of nonlinearity. Several examples \cite{Pugliese_2011,Pugliese_2013,2020EPJC...80..587S,Stuchl_k_2015,Dymnikova_2019,Vieira_2014,2018Stashko,Meliani_2015,Chowdhury_2012} demonstrate occurrence of circular orbit  distributions with several non-connected rings of SCO. This may be of particular interest as observational signs of differences from ordinary black holes, as well as the unusual form of the accretion disk images and/or their radiation properties \cite{Stuchl_k_2014,Schee_2016,2019Stuchlik,paul2019observational,Shahidi_2020,Abdikamalov_2019,Collodel_2021,Shaikh_2019,Gyulchev_2019,Gyulchev_2020}.  

In the present paper, we will look for the effects of nonlinear fields on the SCO distribution (SCOD) around the center. For this purpose, we consider  SFs determined by a sequence of monomial potentials $U(\phi)=\phi^{2n}$,  $n>2$, which have a simple asymptotic behavior $\phi(r)\sim 1/r$ at large distances (analogous to FJNW)\footnote{The cases with $n\le 2$ lead to asymptotics at infinity different from $\sim 1/r$.}. The reason for such choice is that this is the simplest nonlinear generalization of the FJNW  problem. On the other hand, the monomial  potentials are often used in various cosmological problems (see, e.g., 
\cite{Smith_2020,Antusch_2020,Ballardini_2019}) 
We numerically obtain static solutions of the Einstein-SF equations in the case of spherical symmetry and use these results to study SCO with focus on the qualitative features of SCOD, as well as on images of these distributions that can be observed from infinity. Namely, we systematically analyse  the permitted SCO regions.

The paper is organised as follows. In Section \ref{section:BasicRelations} we   write down the basic equations and integrate them numerically.  The use of numerical methods  presupposes that the problem  is well posed. In this regard, we rely on the results of \cite{ZhdSt}, which guarantee that our solutions are regular and they  have no singularities outside the center (in contrast, e.g., to some  special relativistic cases \cite{ZhdSt}). Also, in Appendix \ref{Iterations} we prove that there is a unique solution defined by the boundary conditions at infinity. In Section \ref{section:SCO} we  proceed to test particle  motion in the gravitational field of the configuration.  the "equatorial" plane and present  possible  SCOD. Four qualitatively different SCOD types are introduced, differing in the number of individual SCO rings (subsection \ref{sub_circular orbits}). Here we demonstrate how  the topology of the SCO rings change with $n$  and present maps that define types of SCOD for given configuration parameters. The next subsection \ref{sub_photon trajectories} discusses photon trajectories that are used to build the images of different SCOD. 
The concluding remarks are summarized in Section \ref{sec:discussion} where we discuss  observational signatures of SCOD, which can be used to distinguish them. 
\section{Spherically symmetric static solutions of Einstein relations with scalar field}\label{section:BasicRelations}
The general metric of a static spherically symmetric space-time in the "curvature" coordinates (Schwarzschild-like)  is  
\begin{equation}
\label{metric}
ds^2 = e^{\alpha(r)}dt^2 - e^{\beta(r)}dr^2 - r^2 dO^2,
\end{equation}
where  $dO^2=d\theta^2+(\sin\theta)^2 d\varphi^2$;  radial variable $r> 0$.

We consider one minimally coupled real SF $\phi$ with  Lagrangian density
\begin{equation}\label{lagrangian}
L=\frac{1}{2}\phi,_{\mu}\phi^{,\mu}-V(\phi),
\end{equation}
with 
\begin{equation}
\label{monomial_Self-int}
V(\phi)= |\phi|^{2n} \,, \,\, n>2. 
\end{equation}
where $n$ is not necessarily an integer. More general power law potential can be reduced to (\ref{monomial_Self-int}) by  rescaling of  the variables.  Note that for $V(\phi)=0$ we have the well known FJNW solution. Some results concerning cases  $1\le n<2$ can be found in \cite{stephenson_1962,Asanov1974,St-Zhd_2019UJP,ZhdSt}. 

The Einstein-SF equations are reduced to the following system:  
\be
\label{Ein_1-0}
 	\frac{d}{dr} \left[ r \left(e^{-\beta}-1\right) \right]=-8\pi r^2 [e^{-\beta}\phi'^2/2+ V(\phi)] \,,
\ee
where  $\phi\equiv \phi(r)$,
\be
 	\label{Ein_2-0}
 	r e^{-\beta}\frac{d\alpha}{dr}+e^{-\beta}-1 =8\pi r^2 [e^{-\beta}\phi'^2/2- V(\phi)],
\ee
and
\be
 \label{equation-phi}
\frac{d}{dr}\left[r^2 e^{\frac{\alpha-\beta}{2}}\frac{d\phi}{dr}\right]=r^2 e^{\frac{\alpha+\beta}{2}}V'(\phi)\, .
\ee
Here $\alpha(r),\beta(r)$ are assumed to be $C^1$ functions and $\phi(r)$ is a $C^2$ function for $r>0$. 

We deal with isolated configurations; correspondingly, we impose the asymptotic flatness conditions as follows 
\be\label{flattness}
 \lim\limits_{r\to \infty}\left[ r(e^{\alpha}-1)\right]= \lim\limits_{r\to \infty}\left[ r(e^{-\beta}-1)\right] =-r_g, 
\ee
 where $r_g=2M$ and $M>0$ is the configuration mass; also $\phi(r)\to 0$ for ${r\to\infty}$ and
  \begin{equation} \label{infinity_limit}
\lim\limits_{r\to \infty}
  r^2 \frac{d\phi}{dr}=-Q\,,
  \end{equation}
where parameter $Q$ defines the strength of the scalar field at spatial infinity; we will call it  "scalar charge". Relation (\ref{infinity_limit})  yields 
\begin{equation*}
    \lim\limits_{r\to \infty}
  r\phi(r)=Q\,.
\end{equation*}

The global behavior of the solutions satisfying  (\ref{flattness},\ref{infinity_limit}) has been studied in a more general case of multiple scalar fields \cite{ZhdSt}, where a proof of regularity of solutions on open interval $(0,\infty)$ is given. Asymptotic properties for $r\to\infty$ have been derived in   \cite{ZhdSt}  for a particular case, under the assumption that the solutions can be expanded  in powers of $1/r$. In Appendix \ref{Iterations} of the present paper we provide a more rigorous  analysis  by means of an iteration procedure. We prove  that there is a solution  of equations (\ref{Ein_1-0}, \ref{Ein_2-0}, \ref{equation-phi}) for $r\ge r_{\rm in}$, where   $r_{\rm in}$ is large enough, with the conditions (\ref{infinity_limit}); this  solution is uniquely defined by parameters  $M,Q$. 
The first iterations of this procedure yield asymptotic relations for large $r$ as follows: 
\be
\label{asymptotics_inf_metric}
\phi(r)=\frac{Q}{r}\left[1+\frac{r_g}{2r}+\frac{n|Q|^{2n-2}}{(n-2)(2n-3)r^{2n-4}} \right]+O\left[\frac{\mu(r)}{r^2}\right] \,,\quad (n>2),
\ee
\be
\label{asymptotics_inf_metric1}
e^{\alpha}=\left(1-\frac{r_g}{r}\right)\left[1+O\left(\frac{\mu(r)}{r^2}\right)\right] ,~ e^{- \beta}= \left(1-\frac{r_g}{r}\right)\left[1+\frac{4\pi Q^2}{r^2}+O\left(\frac{\mu(r)}{r^{2}}\right)\right]\,,
\ee
where $\mu(r)= 1/r$ for $n\ge 3$ and $\mu(r)= 1/r^{2n-4}$ for $2<n<3$.
Note that in general case  (\ref{asymptotics_inf_metric},\ref{asymptotics_inf_metric1}) can contain non-integer powers of $1/r$.

Asymptotics of the metric near the center can be found in \cite{ZhdSt}: $\alpha(r)\sim{(\eta-1)}{\ln r},\quad
\beta(r)\sim{(\eta+1)}{\ln r}$.   
Parameter $\eta$  characterizes the strength of the singularity. There is a critical point $\eta=3$ that separates two  types of the singularity with different behavior of the null geodesics (see below).

\begin{figure*}[t!]
    \centering
\includegraphics[width=80mm]{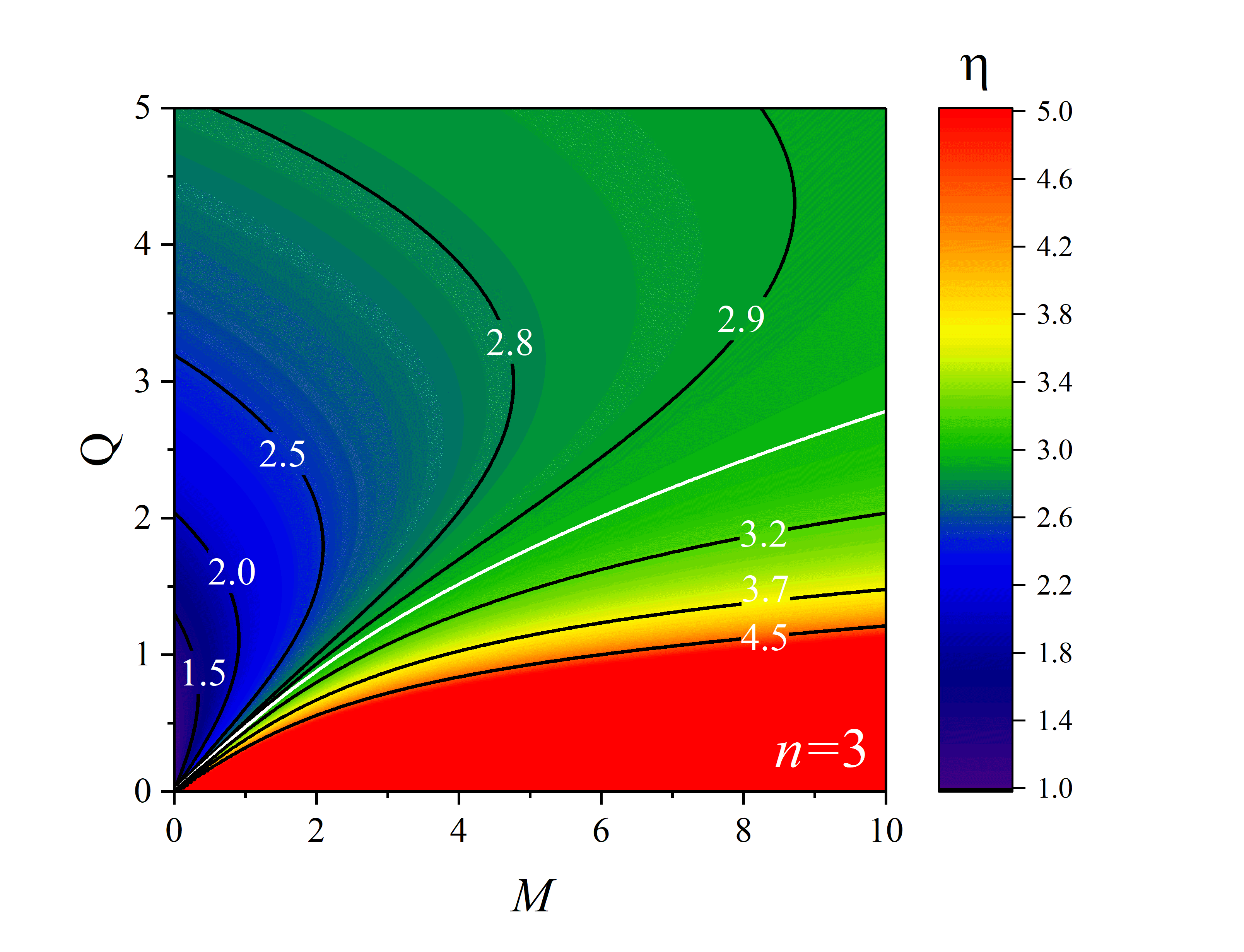}
\includegraphics[width=80mm]{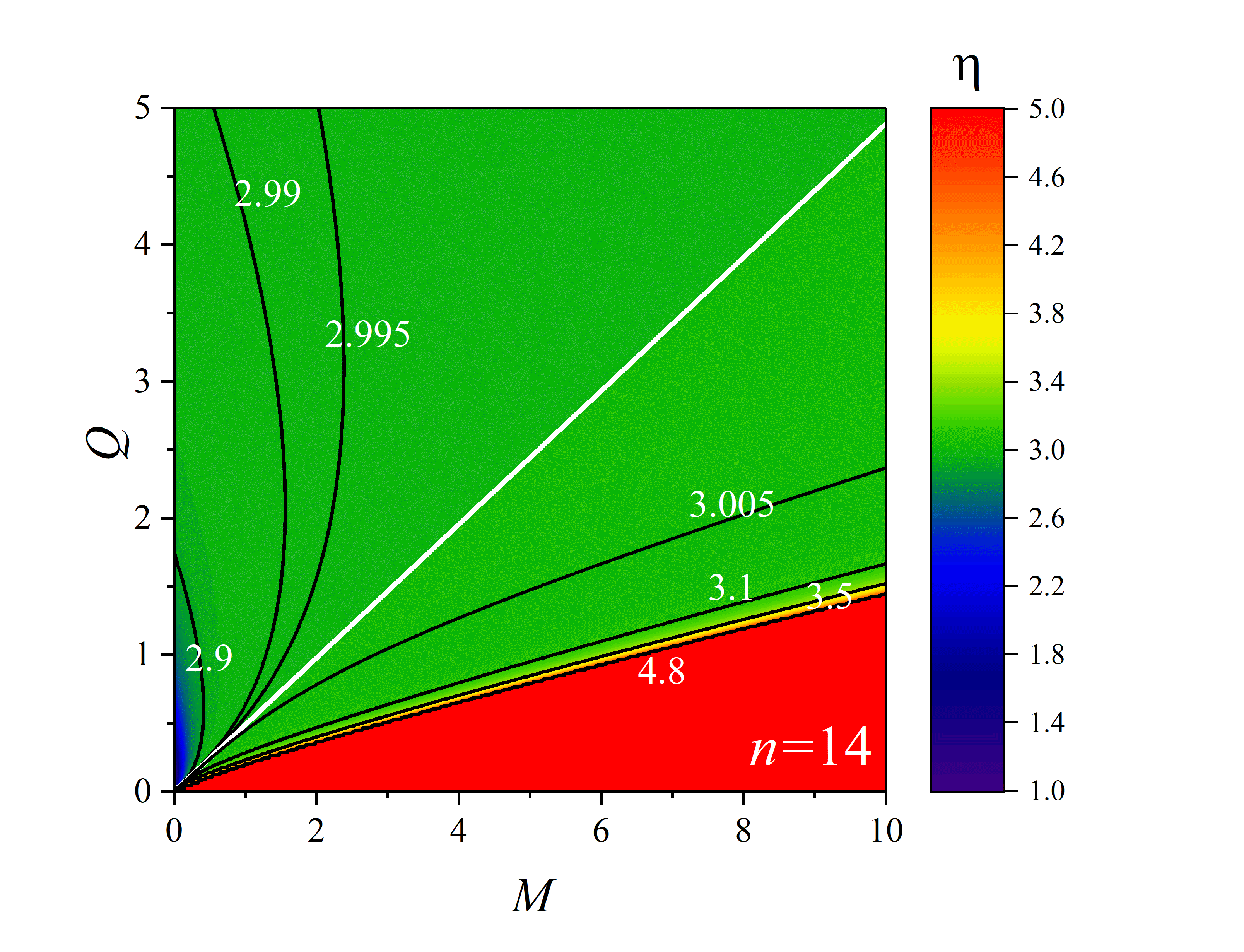}
    \caption{Contours of the equal $\eta$ as function of $M,Q$ for $n=3$ (left panel) and $n=14$ (right panel). Solid white line corresponds to  critical value $\eta=3.$ If $Q\rightarrow{0}$,  then $\eta\rightarrow{\infty}$.}
    \label{fig2}
\end{figure*}

We performed a detailed numerical investigation of the problem (\ref{Ein_1-0}--\ref{infinity_limit})  for  $n\le 40$, $M\le 60$, $|Q|\le 60$. To find the solutions  for $r\leq r_{\rm in}$, we  proceed numerically starting at  sufficiently large initial radius $r_{\rm in}$ (up to $10^5$) to use  initial conditions in accordance with  asymptotic relations (\ref{asymptotics_inf_metric}, \ref{asymptotics_inf_metric1}). We integrate backwards from higher to lower  values of $r\in (0,r_{\rm in}]$.  
The qualitative properties of the metric coefficients and scalar field are rather similar for different values of parameters $(Q,M,n)$:   ${\alpha(r)}$ is monotonically increasing function bounded from above by 1 for $r\rightarrow{\infty}$ and   ${\beta(r)}$ has a maximum at some point $r=r_{\rm max}(Q)$.  If $Q$ increases, then $r_{\rm max}(Q)$ is shifted to larger values and the maximum becomes less pronounced. 
The scalar field is always  a monotonically decreasing function and $\phi\rightarrow{0}$  as $r\rightarrow{\infty}$.
For  large $n$ and fixed $M,Q$, the solutions approach the FJNW curves, except for a small region near the singularity. 
 An important point is the dependence of $\eta$ upon the parameters of the configuration at spatial infinity.  Qualitatively, dependencies $\eta(M,Q)$ for different $n$ are rather similar; Fig.  \ref{fig2} shows the examples  for $n=3$ and $n=14$.  \\
\\
\FloatBarrier

\section{Test particle motion}\label{section:SCO}
\subsection{General relations}\label{section:SCOa}
This Section deals with trajectories of  the test particles in the space-time corresponding to the solutions of the problem (\ref{Ein_1-0}--\ref{infinity_limit}).  
The equations of the test particle motion follow from formal Lagrangian 
\begin{equation}
  L=g_{\mu\nu}\frac{dx^{\mu}}{d\tau}\frac{dx^{\nu}}{d\tau},
\end{equation}
where $\tau$ is a canonical parameter.  Standard procedure involves the first integrals for trajectories in the equatorial plane ($\theta=\pi/2$): 
 \begin{equation}
 \label{geodesics_1}
 e^{\alpha}\left(\frac{dt}{d\tau}\right)^2 - e^{\beta}\left(\frac{dr}{d\tau}\right)^2  - r^2\left(\frac{d\varphi}{d\tau}\right)^2=S\,,
 \end{equation}
  \begin{equation}
 \label{geodesics_2}
 e^{\alpha}\left(\frac{dt}{d\tau}\right)=E, \quad  r^2\left(\frac{d\varphi}{d\tau}\right)=L\,,
 \end{equation}
 where $S=0$ in case of null trajectories and $S=1$ for the test particles with the non-zero mass; $L,\,E$ are the integrals  of motion.
 This yields
 \begin{equation}
 \label{1D particle}
 e^{\alpha+\beta}\left(\frac{dr}{d\tau}\right)^2=E^2-
 U_{\rm eff}(r,L,S)  \,,
 \end{equation}
 where $ U_{\rm eff}(r,L,S)=e^{\alpha}\left(S+{L^2}/{r^2}\right)$.

Thus, we the problem is reduced to investigation of the  one-dimensional particle motion in the field of  effective potential $U_{\rm eff}$. The asymptotic behavior of the potential  depends on asymptotics  (\ref{asymptotics_inf_metric1}) at infinity   and  those near singularity (see \cite{ZhdSt}). For $L\neq 0$  we have 
\be \label{U_eff-asymptotics}
U_{\rm eff}(r\rightarrow{0})\sim r^{\eta-3}\,,\quad
U_{\rm eff}(r\rightarrow{\infty})\rightarrow{S}\,.
\ee  
Thus for $\eta<3$ we have  $U_{\rm eff}\rightarrow{\infty}$ at $r\rightarrow{0}$, i.e.  there is an infinite potential barrier that will reflect falling particles. For  $\eta>3$, $U_{\rm eff}\rightarrow{0}$ at $r\rightarrow{0}$, the particles can approach the center and there is a maximum at some $r_u>0$, which  is the radius of an unstable circular orbit.

\subsection{Circular orbits}\label{sub_circular orbits}
We consider  the  Page-Thorne model of the geometrically thin accretion disk (AD)  \cite{Page_Thorne}, where the averaged motion of accretion matter is approximated  by  means of SCO around gravitating center. 
A circular orbit is stable if the right hand side of (\ref{1D particle}) is zero and $dU'_{eff}/dr$ changes its sign at this point. 
For the congruence of circular orbits with different radii in  equatorial plane  we get  dependencies of the specific energy and  the specific angular momentum, and the angular velocity   $\Omega=d\varphi/dt$ upon radius $r$ as follows
\be
\label{tilde_E_L_Omega(r)}
  \tilde E^2(r)=\frac{2e^{\alpha(r)}}{2-r\alpha'(r)},~~ \tilde L^2(r)=\frac{r^3\alpha'(r)}{2-r\alpha'(r)},~~\Omega^2(r)=\frac{\alpha'(r)e^{\alpha(r)}}{2r}
\ee

\begin{figure}[h!]
    \centering
\includegraphics[width=80mm]{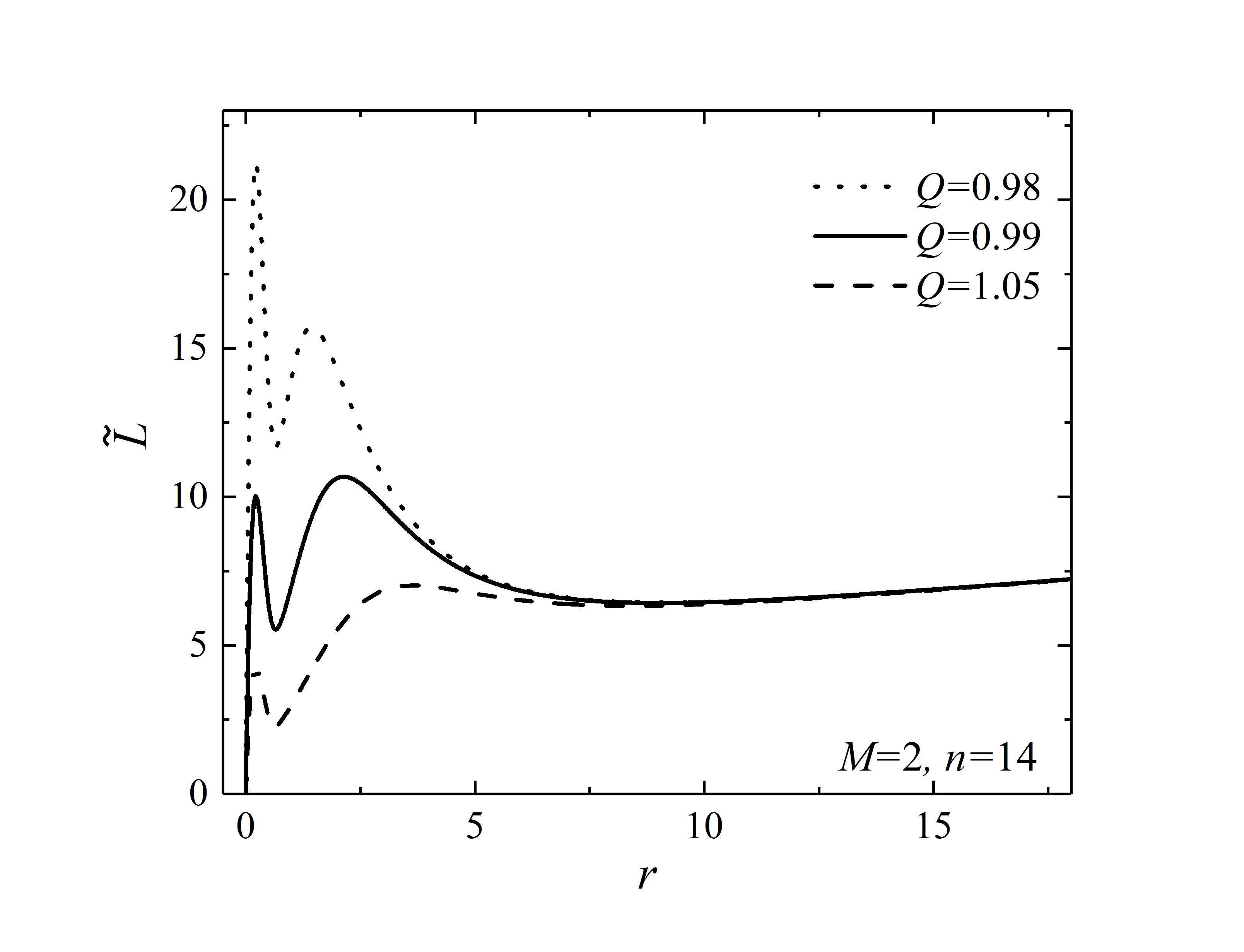}
\includegraphics[width=80mm]{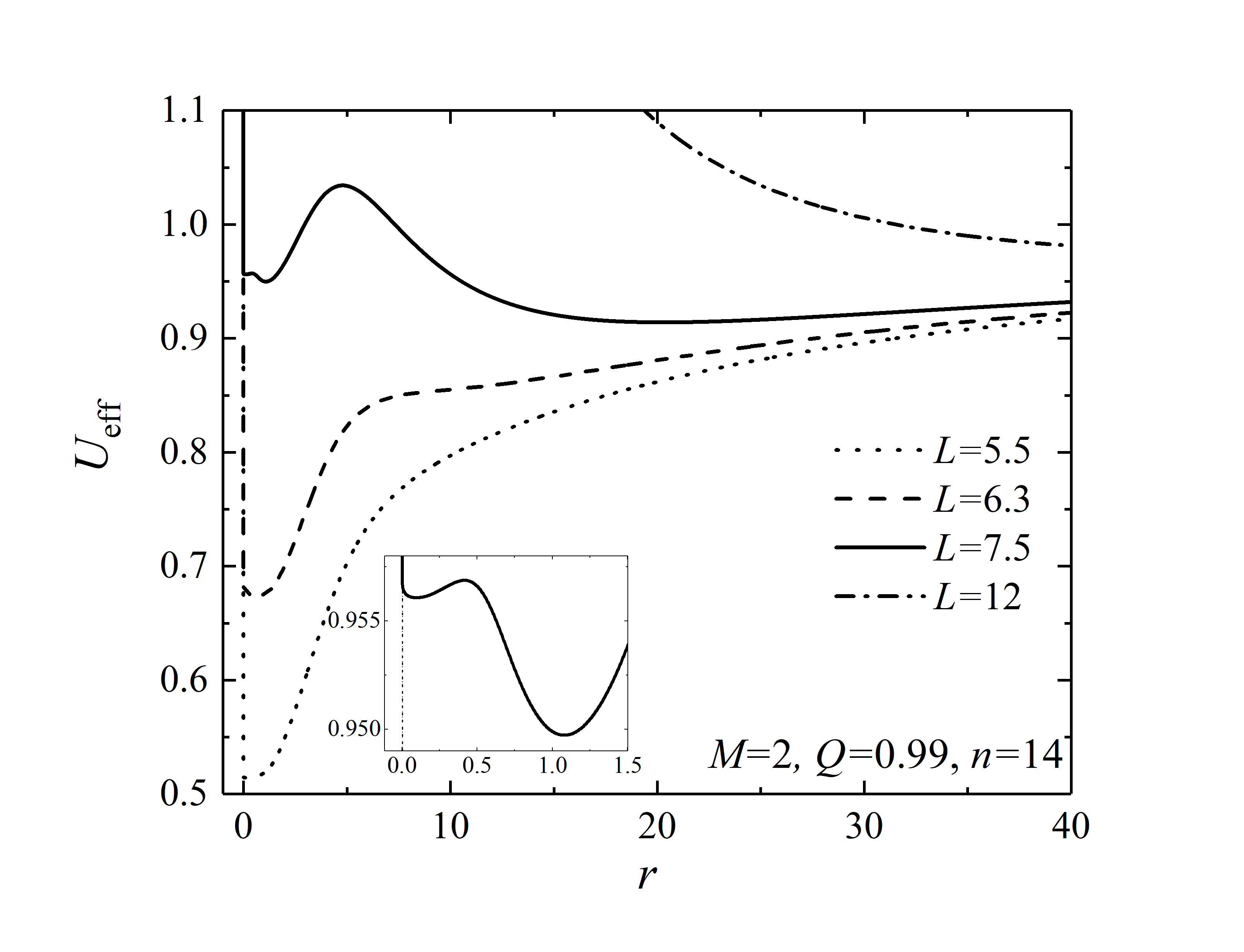}
\caption{Typical dependencies $\tilde L(r)$ and $U_{\rm eff}(r)$ in case of  {\it S3} type; $\eta<3$ and $U_{\rm eff}(r)$ is infinite for $r\to0$. Left panel shows that there can exist several circular orbits with the same $\tilde L(r)=L$. Roots of this equation  correspond to SCO if they belong to the intervals where  $\tilde L'(r)>0$.  Right panel shows   effective potential with three minima ($S3$ type). Three different SCO rings can exist near $L\sim 6.3$ and about fixed  $M,Q,n$ indicated in the figure.}
    \label{fig:case3}
\end{figure}

We use a semi-analytical  method of our works \cite{2018Stashko,Stashko_Zhdanov_2019a} to study bifurcations associated with the appearance and disappearance of the minima of $U_{\rm eff}$. Essentially this is connected with investigation of joint  conditions $U'_{\rm eff}=0$ and $U''_{\rm eff}=0$, which allow us to exclude $L$; this leads to a necessary condition $F(r)=0$, where
\begin{equation} \label{bifurcation_condition}
  F(r)= \tilde L^2(r) [r^2\alpha''(r)-2r\alpha'(r)+6]+r^4\alpha''(r).
\end{equation}
Using equation (\ref{bifurcation_condition}), we get the bifurcation values $r_b, \, L^2_b=\tilde L^2(r_b)$ and $E^2_b=\tilde E^2(r_b)$ under conditions  that  $\tilde E^2(r_b)>0$ and $\tilde L^2(r_b)>0$. 
We tested our results  by considering the explicit  forms of $U_{\rm eff}$ near the bifurcations as in the example in Fig. \ref{fig:case3}. The right panel of 
Fig. \ref{fig:case3}  illustrates occurrence of three minima (which is not observed in case of FJNW solution); it  shows how the shape of the potential changes with increasing the  angular momentum, and as a result, the third  minimum appears in the vicinity of the singularity. Note that the bifurcation radius  (boundary radius), signalizing emergence or disappearance of some minimum of  $U_{\rm eff}$, indicates the boundary of some SCO ring for a given configuration. 


 The above reasoning allowed us to obtain the boundaries of rings containing SCO, which are  separated by regions where circular orbits are either unstable or do not exist.
Figs. \ref{fig:example}, \ref{fig:ISCOr1}, \ref{fig:ISCOr2} below present examples of bifurcation curves, which define these boundaries. 

 Here we describe main results of our  numerical investigation of the bifurcation radii $r_b\equiv r_{i(T)}$, where index $T$ denotes various SCOD type explained below. We show that at least four SCOD types  are possible in our problem. 
 
 The next 3 types deal with the case of $\eta<3$ (effective potentials are unbounded)
\begin{itemize}
    \item {\bf {\it S1}}:  $\tilde L(r)$ is monotonically increasing function. As a result we have one connected domain of SCO  with radii   $r\in(0,\infty)$. SCO fill all the space.
    \item {\bf {\it S2}}:  The effective potentials $U_{\rm eff}$ have two minima, which correspond to the  SCO radii.   As a result we have inner disk with SCO radii $r\in(0,r_{1(2)})$ and outer disk with $r\in(r_{2(2)},\infty)$. Hereinafter, the index in parenthesis denotes the type. 
    
Note that types {\it U1, S1, S2} exist in the FJNW case.
    
    \item {\bf{\it S3}}:  We found a new type of the SCO distribution that does not exist in the FJNW case. The effective potential $U_{\rm eff}$ can have three possible minima and two maxima corresponding to three  SCO regions $r\in(0,r_{1(3)})$, $(r_{2(3)},r_{3(3)})$,$(r_{4(3)},\infty)$. These regions are separated by two rings of the unstable orbits with $r\in (r_{1(3)},r_{2(3)})$, $(r_{3(3)},r_{4(3)})$. Fig.  \ref{fig:case3} shows typical dependencies of $U_{\rm eff}(r,L)$ and $\tilde L(r)$. The emergence of SCO with small radii $r<r_{1(3)})$ occurs in a neighborhood  of a change in the shape of $U_{\rm eff}$ at $\eta$ values close to 3, where the type of the singularity  changes. 
    \end{itemize}


If $\eta\geq3$, then effective potential  has a  maximum $U_{\rm eff}(r_{\rm max},L)$ for any $L$. There is a region where either there are no circular orbits at all, or they are unstable.  
 
 Here we have only one type:
\begin{itemize}
    \item {\bf {\it   U1}} (Schwarzshild-like SCO distribution):  effective potential $U_{\rm eff}$ is bounded from above;  $U_{\rm eff}\rightarrow{0}$ for $r\rightarrow{0}$  and only one minimum can exist. We have one unstable and one stable regions of SCO; the latter for $r\in(r_{1(U)}, \infty)$ with lower boundary $r_{1(U)}$.
\end{itemize}


Fig. \ref{fig:example}   show two illustrative examples of bifurcation radii $r_b(Q)\equiv r_{i(T)}(Q)$, $T$ means type of SCOD,  as a functions of $Q$, which realize different SCOD types. 
Fig. \ref{fig:ISCOr1} shows essentially the same, but with more complete picture of the bifurcation radii, which illustrates how the curves transform for larger $n$ values.
Analogous dependencies are shown on Fig. \ref{fig:ISCOr2} as functions on $M$. Figs.  \ref{fig:ISCOr1} and \ref{fig:ISCOr2} show how the shape of the curves changes when $n$ passes some critical values.

\FloatBarrier
\begin{figure}[h!]
 \includegraphics[width=80mm]{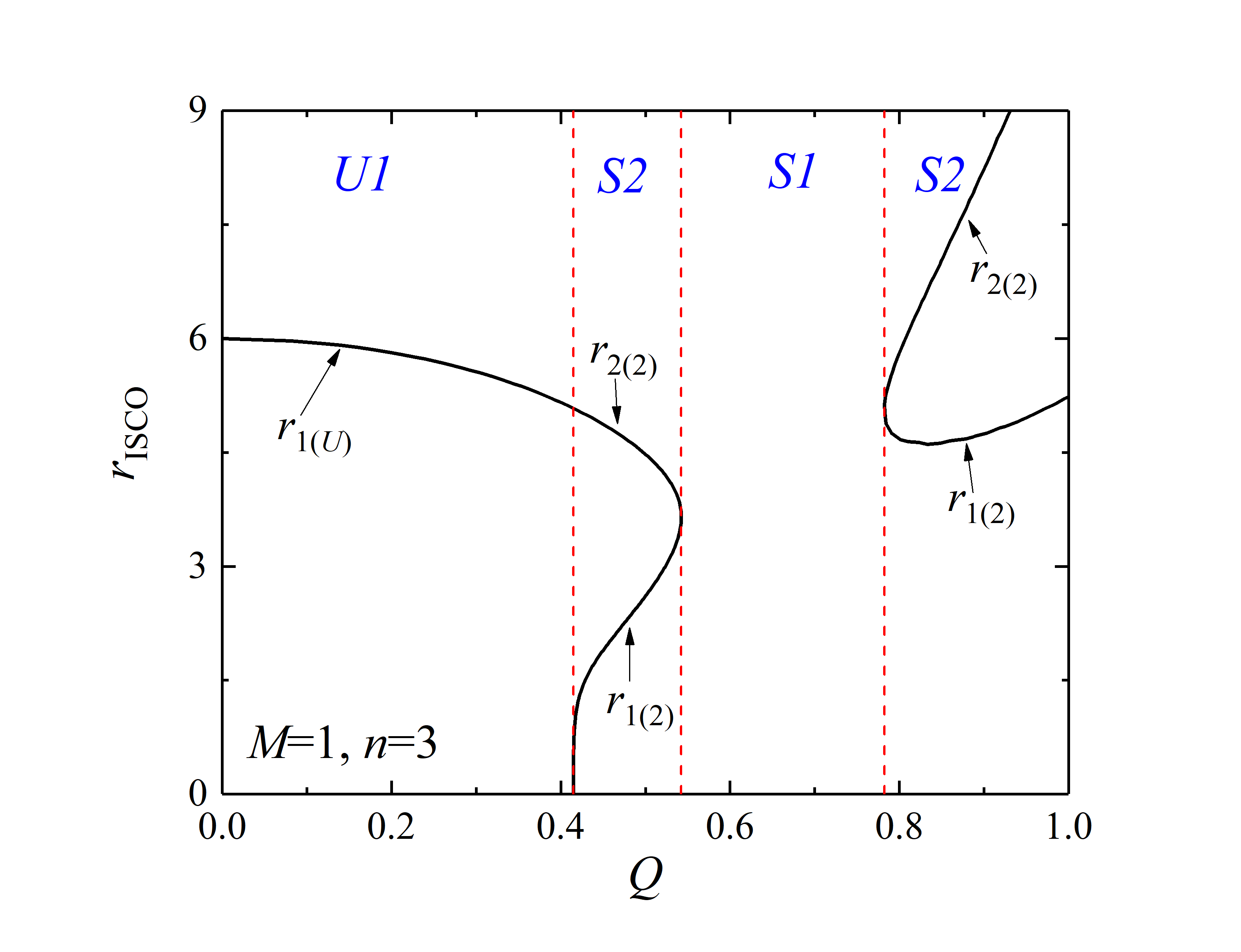}
\includegraphics[width=80mm]{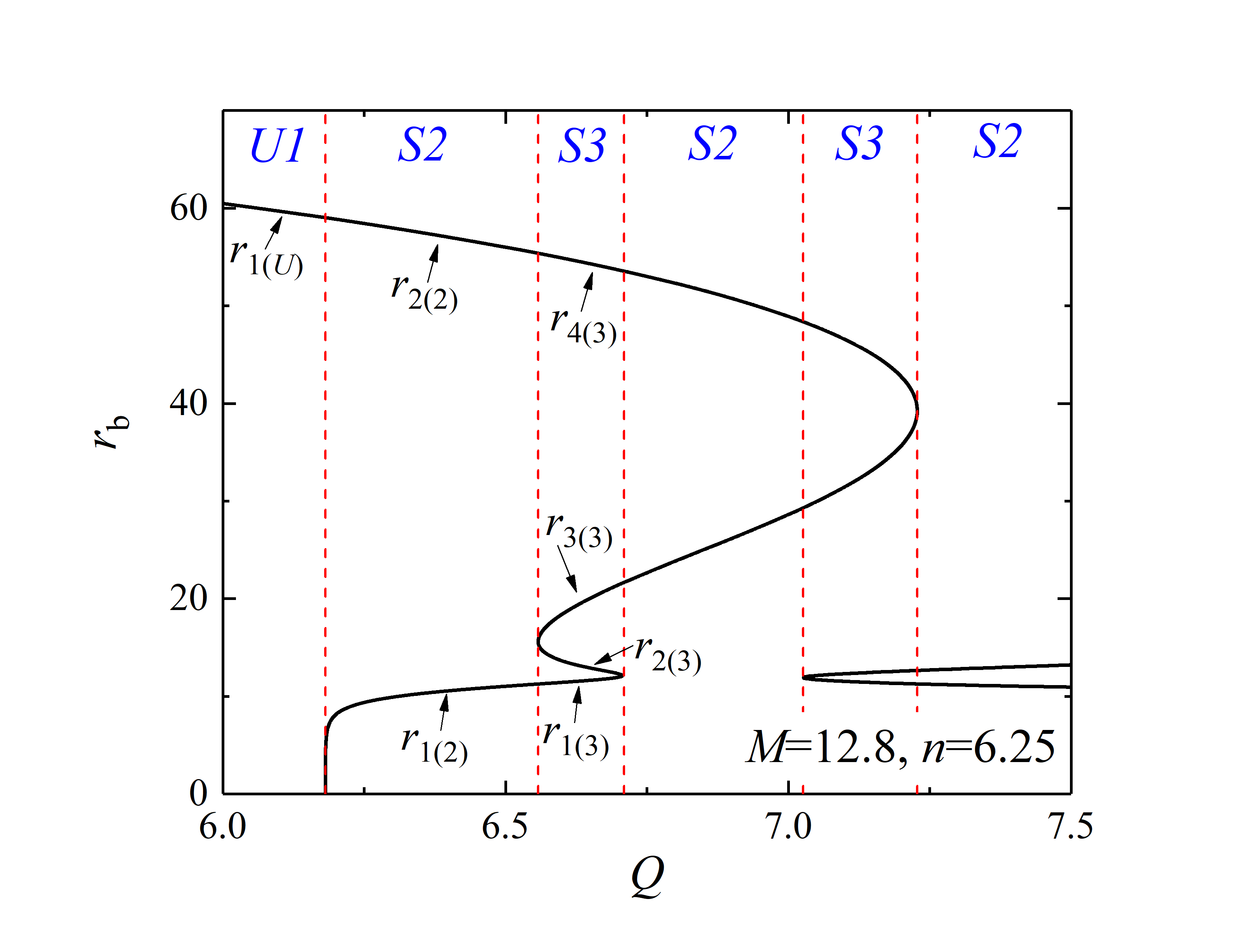}
\caption{Boundary radii of SCO regions as a function of $Q$ for some fixed $M,n$. Vertical dashed lines separate areas of different SCOD types. In the left panel, we have the {\it S1} area between two branches of the bifurcation curve; here    arbitrary SCO radii are allowed. The other allowed region of radii are according to definition of {\it U1} and {\it S2}. The right panel shows the example with larger $M,Q,n$, when another type is present, namely {\it S3}; corresponding parameters fall into the yellow region on Figs. \ref{fig:nQ1},\ref{fig:nQ2}}
\label{fig:example}
\end{figure}

 For every $M$ there are two branches of the  curve  $r_b(Q)$ (unlike the FJNW case, where there is a single branch exists) and there is a sequence of  values  $n^{*}_1<n^{*}_2<n^{*}_3<n^{*}_4<n^{*}_5$, which has the following properties.  
\begin{itemize}
\item The first and the last values denote critical numbers  $n^{*}_1$, $n^*_5$, which correspond to a reshaping (reconnection) of two branches.
\item   For $2<n<n^*_1$, both branches are unbounded (like two solid curves on the left panel of Fig. \ref{fig:ISCOr1}).
After the reshaping, for $n\in(n^*_1,n^*_4)$ with some $n^*_4>n^*_1$, the left branch comes close to the FJNW curve (practically merges), whereas the right one is unbounded (see dashed and dotted curves on the left panel of Fig. \ref{fig:ISCOr1}).
\item For  $n^*_1<n<n^*_2$
the right branch moves away to the right and then, for $n>n^*_2$,  it starts moving to the left. It is important to note that for moderate $n\lesssim 7$,  the sizes of unstable regions can be noticeably larger than in  the FJNW case. 
\item At  $n=n^*_3$, an additional wedge-shaped feature in the left branch appears. This is difficult to show in the right panel of Fig. \ref{fig:ISCOr1} for $M=1$, but this  is well seen for larger $M,Q$ in the right panel of Fig. \ref{fig:example}: the corresponding feature is formed by  sections $r_{1(3)}$ and $r_{2(3)}$ within the area of $S3$ type. This corresponds to  "Pinocchio's nose" in the right panel of Fig. \ref{fig:ISCOr2}. 
\item For $n\in (n^*_4,n^*_5)$ the right branch returns closer enough to the left one and a new $S3$ area appears. See the right panel of Fig. \ref{fig:example}, where there is an additional small $S3$ area  due to the sharp wedge for $r_b\sim 10,\, Q\gtrsim 7$. 
\item At $n=n^*_5$ the tips of the two wedges (see the right panel of Fig. \ref{fig:example}) touch each other.  The new reshaping of the bifurcation curve occurs, after which the branches reconnect forming a structure represented by the solid curve on the right panel of  Fig.\ref{fig:ISCOr1}. For large enough $n$, the lower branch tends to the abscissa axis. 
\end{itemize}

\begin{figure}[h!]
    \centering
\includegraphics[width=70mm]{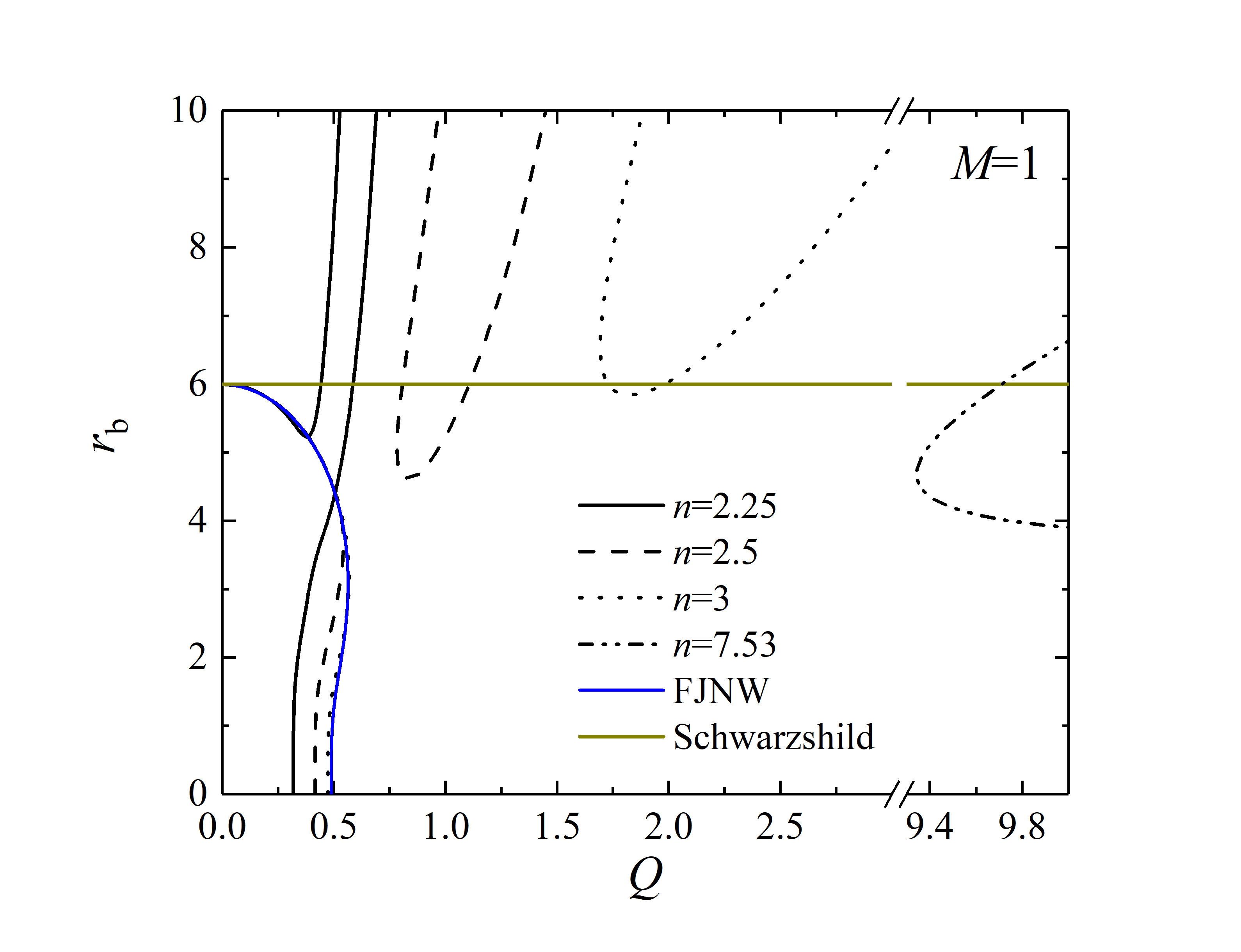}
\includegraphics[width=70mm]{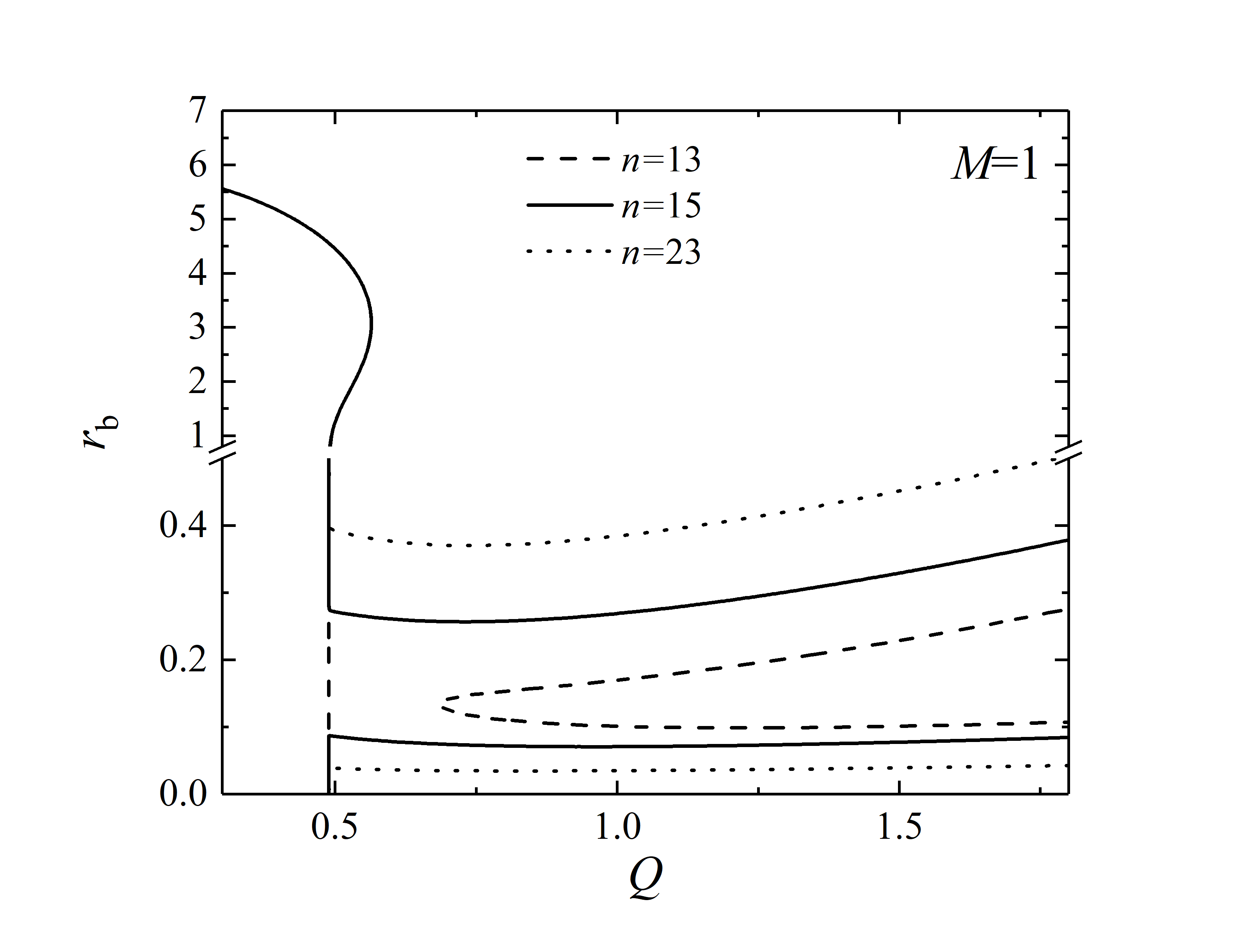}
    \caption{Boundary radii $r_b$ of SCO regions as a functions of $Q$ for $M=1$ and several fixed values of $n$. The continuous blue  line on the left panel represents the FJNW solution and the orange straight line corresponds to the  Schwarzschild BH case ($r_{\rm ISCO}=6M$). The first reshaping occures at  $n^*_1\approx 2.38$; then the right branch moves to the right and  then starts backtracking to the left branch  when $n>n^*_2\approx 7.53$. We found  $n^*_3\approx 13.12$ and $n^*_4\approx 13.13$. For ($n\geq n^*_5\approx13.15$, the form of the branches is represented by solid ($n=15)$  and dotted ($n=23)$ lines on the right panel; here we also have the $S3$ type (four-valued function $r_b(Q)$ on some interval of $Q$).}
    \label{fig:ISCOr1}
\end{figure}

\begin{figure}[h!]
    \centering
\includegraphics[width=70mm]{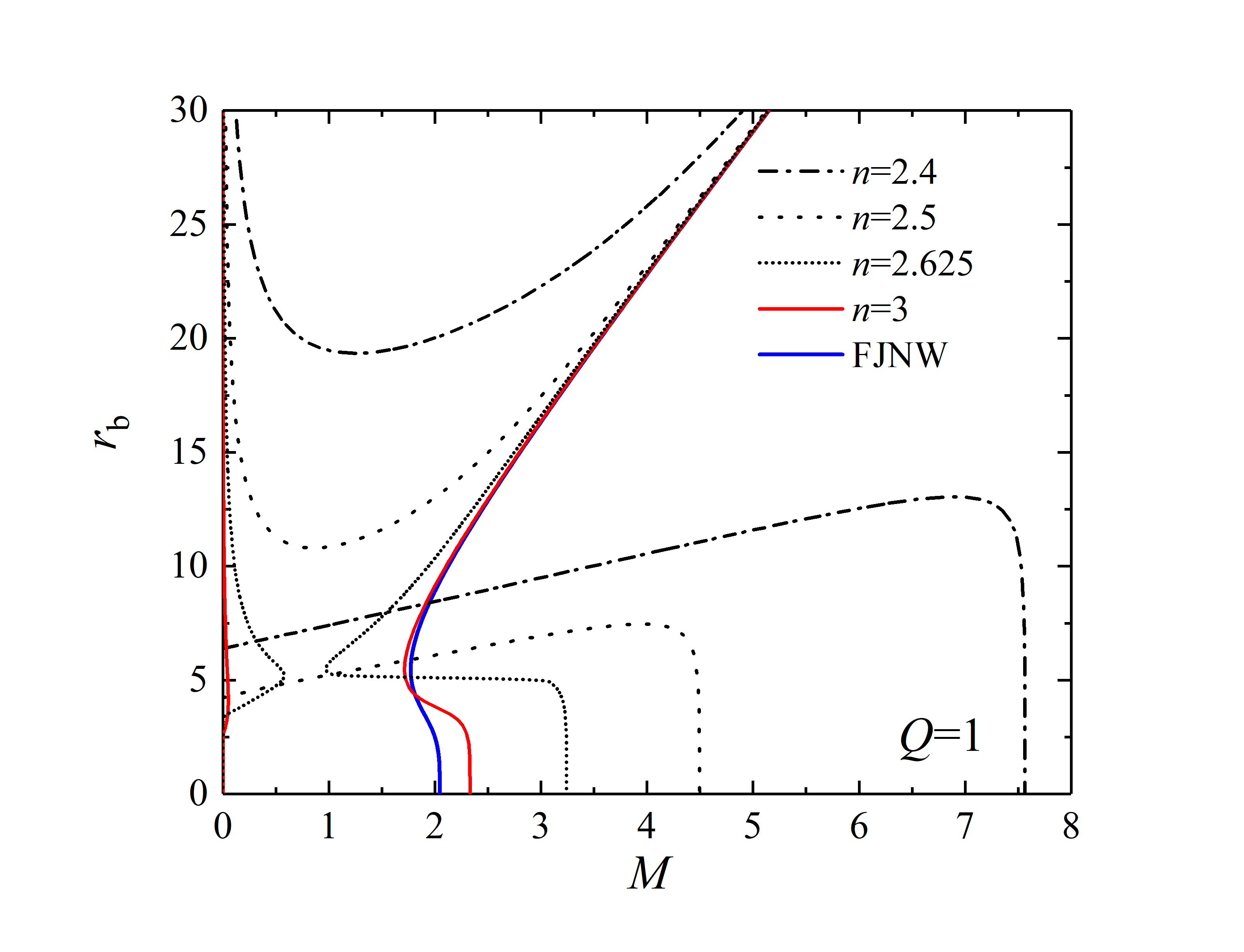}
\includegraphics[width=70mm]{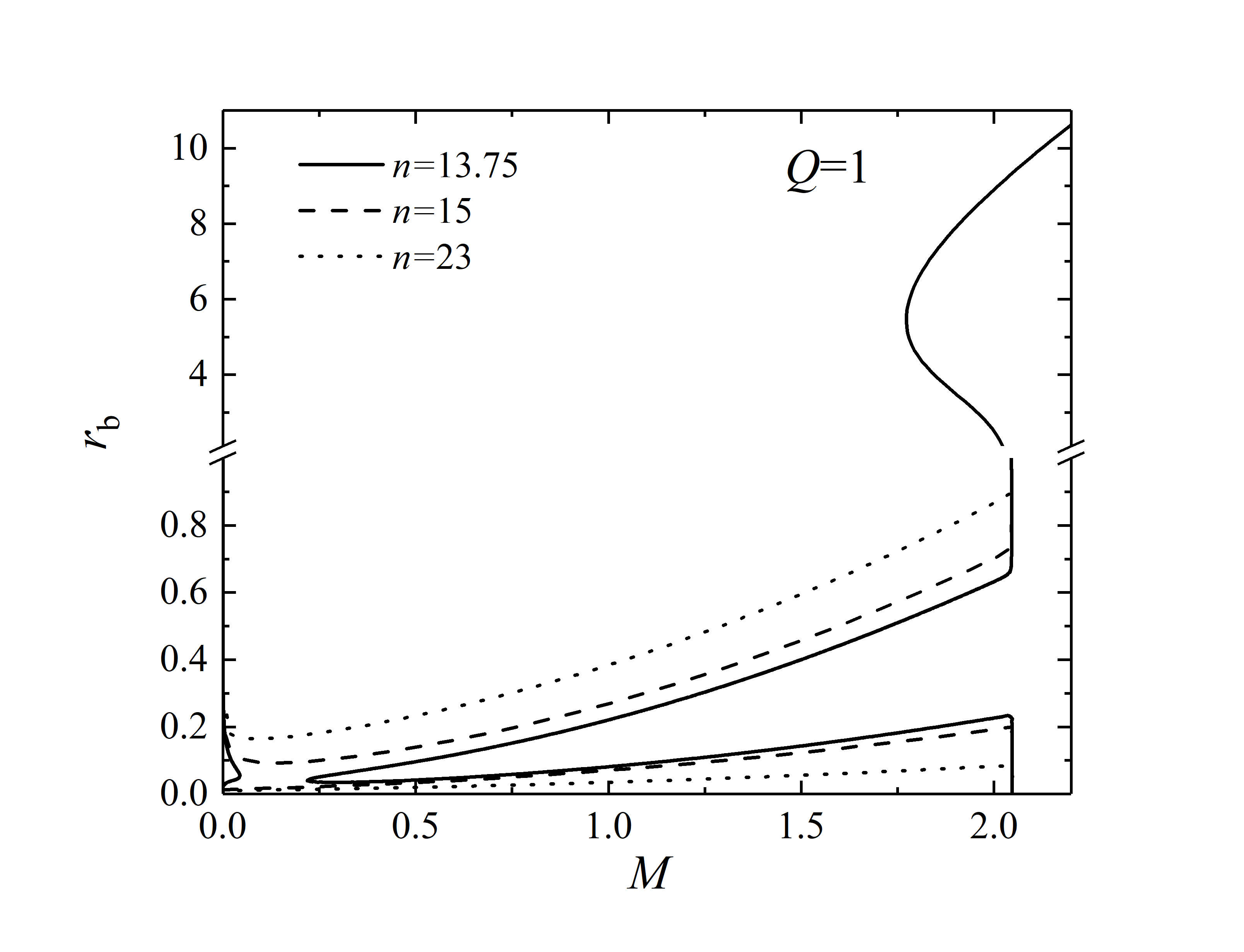}
    \caption{Boundary radii $r_b$ of SCO regions as a functions of $M$ for $Q=1$ and several fixed values of $n$.
    The first reconnection of two branches occurs at $n\approx2.62$; dash-dotted ($n=2.4$) and dotted ($n=2.5$) lines show typical form of branches before the reshaping (left panel), the  lines with $n=2.625$  and $n=3$ show the typical forms after the reshaping. The right panel demonstrates the branches around the second reconnection near $n\approx 13.86$.}
    \label{fig:ISCOr2}
\end{figure}

\begin{figure}
    \centering
\includegraphics[width=70mm]{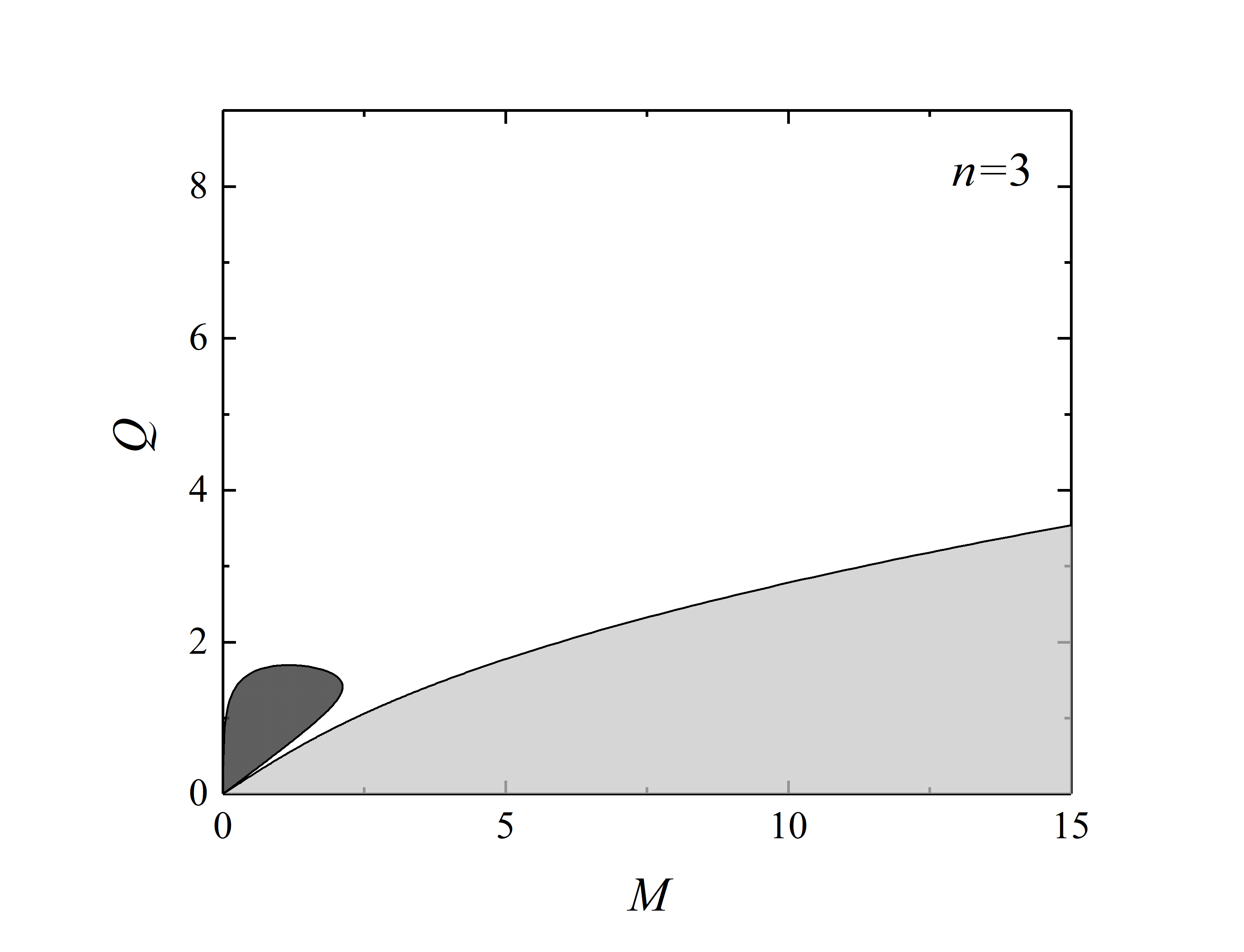}
\includegraphics[width=70mm]{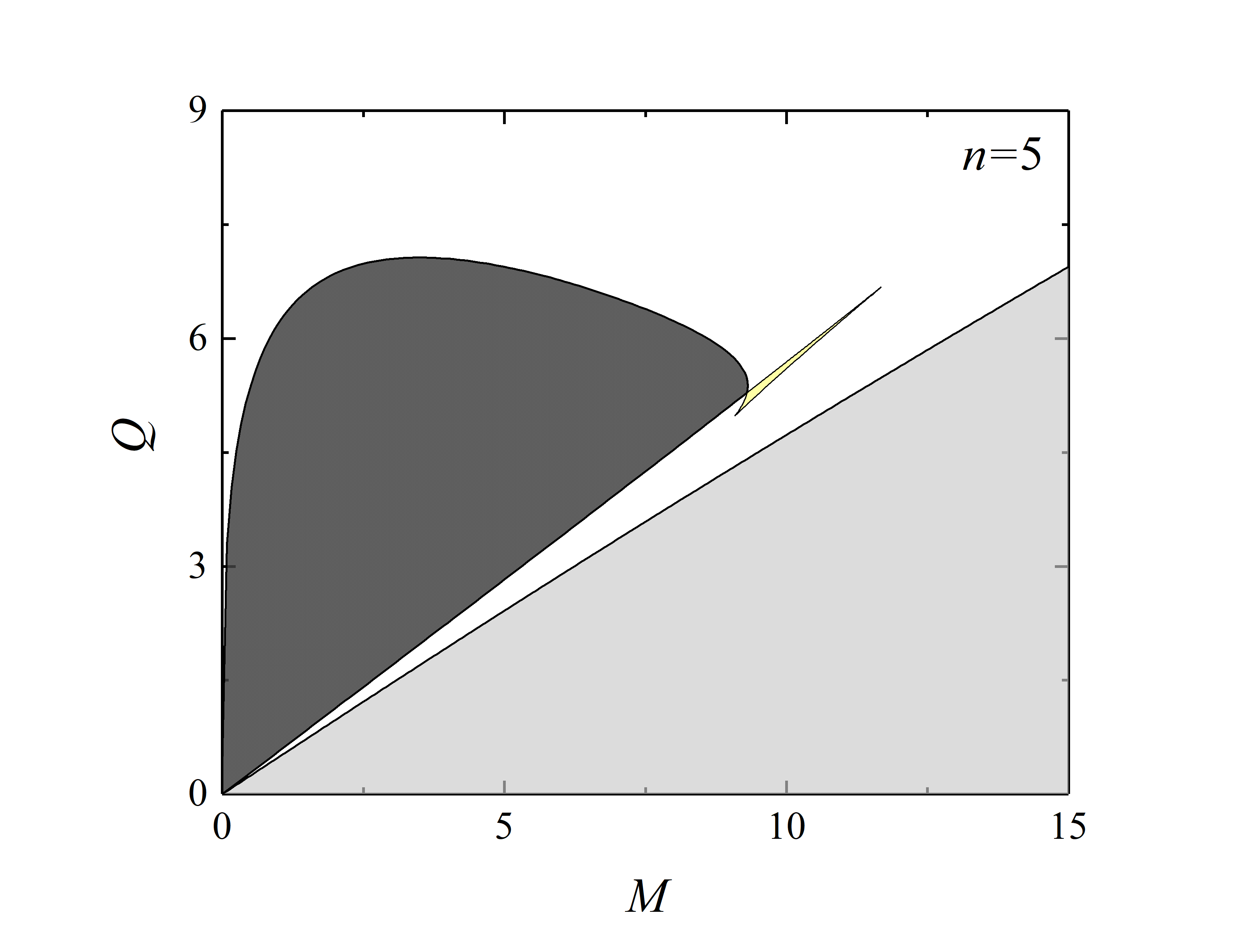}
  \caption{Domains of parameters on $(M,Q)$ plane for different values of $n.$ Here and on figures below white colour denotes the $S2$ type, yellow-$S3$, light gray-$U1$, dark gray $S1$. For  values of $n\sim 7 $ the $S1$ area grows.}
    \label{fig:MQ1}
\end{figure}

\begin{figure}
\includegraphics[width=70mm]{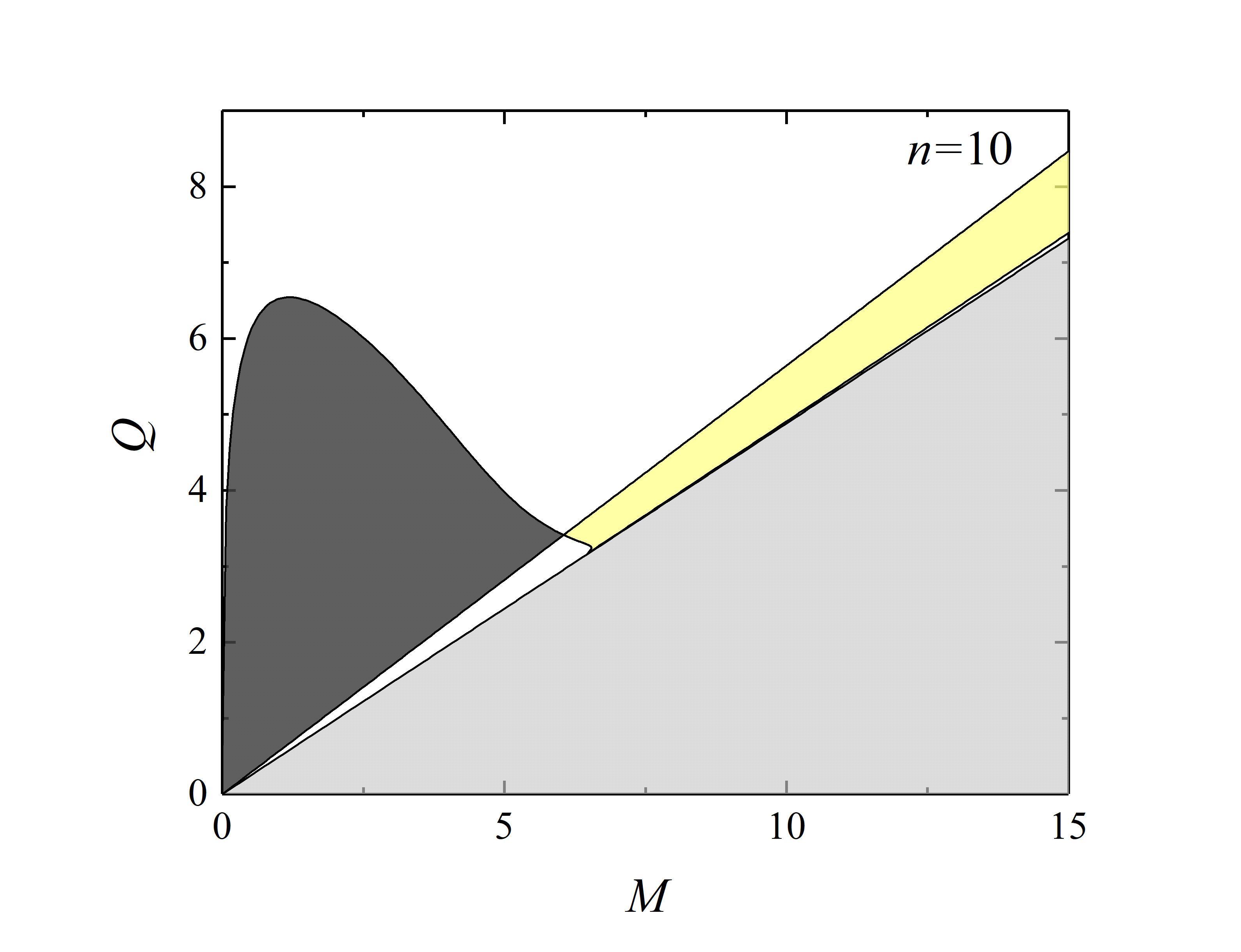}
\includegraphics[width=70mm]{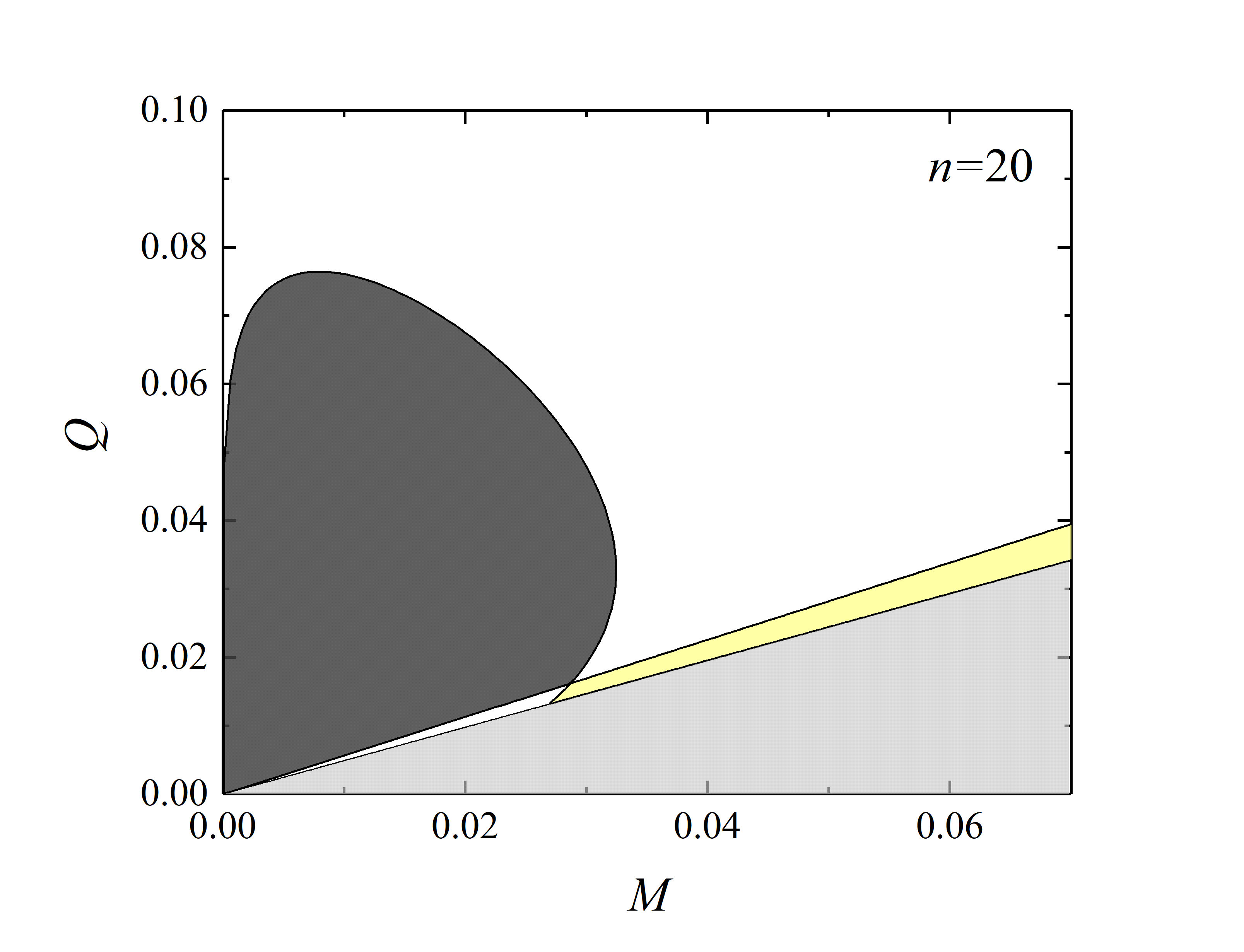}
    \caption{For larger $n\gtrsim 7$ the black region ($S1$) shrinks to the origin and $S2$ dominates.  }
    \label{fig:MQ}
\end{figure}

We have determined the number of rings with stable and unstable circular orbits and built maps of possible SCODs in the $M-Q$ plane for different $n$. Each point on every map (see Figs. \ref{fig:MQ1} -- \ref{fig:MQ}) shows the type of SCOD. 

For all $M$ and $n$ there are  $S2$ and $U1$ types  with appropriate  $Q$. We note that the "thickness" of the ring in the $S2$ case can be considerably larger than in the FJNW case. 
The size of $S1$ area changes non-monotonically: it increases up to $n\sim 7$ and then decreases.

We discovered that the $S3$ region emerges for  $n\sim 4.320$ (Fig. \ref{fig:MQ1}, right); for  larger $n$ this region  becomes to appear for smaller  $M$  (Fig. \ref{fig:MQ}, right).
 

\begin{figure}
    \centering
\includegraphics[width=70mm]{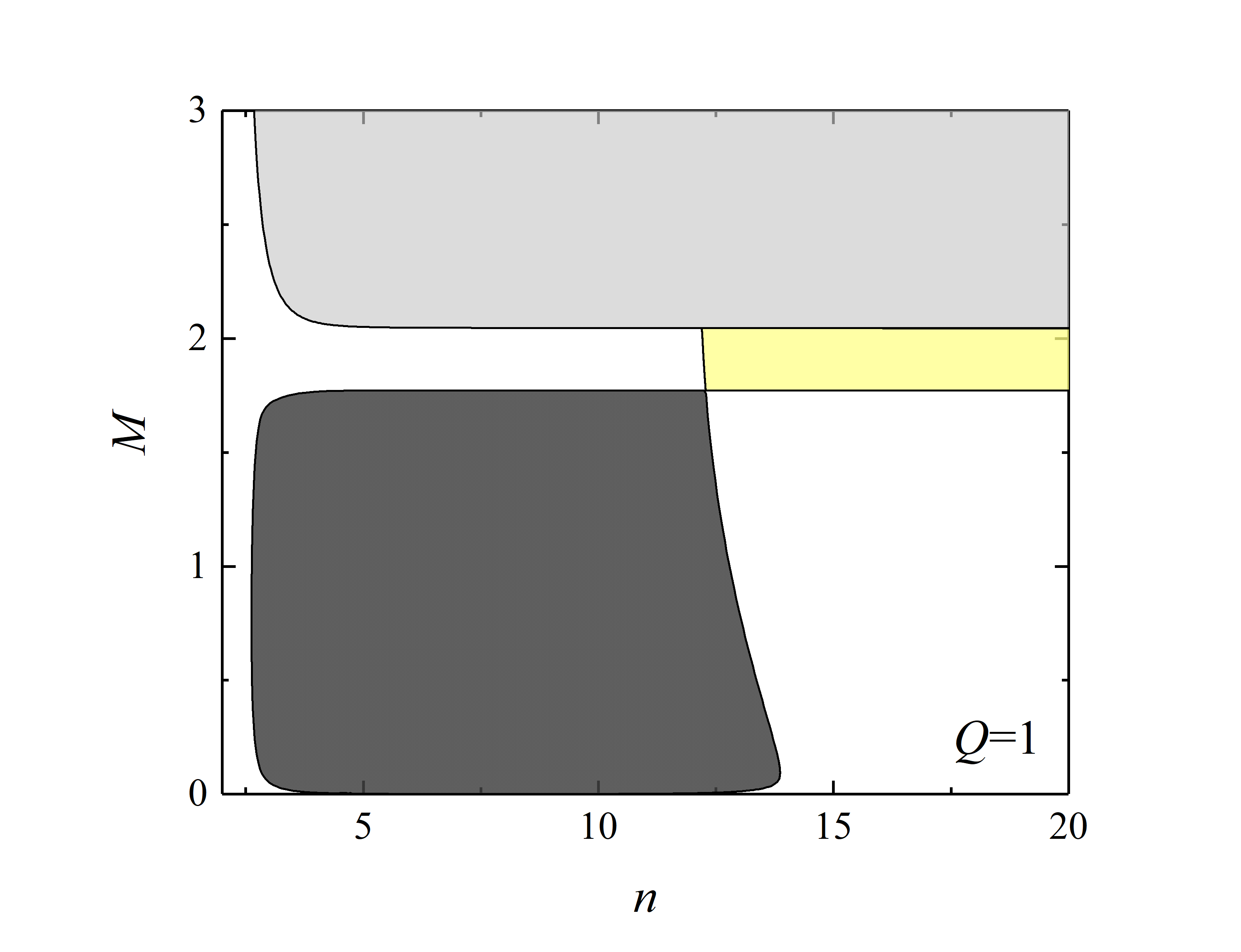}
\includegraphics[width=70mm]{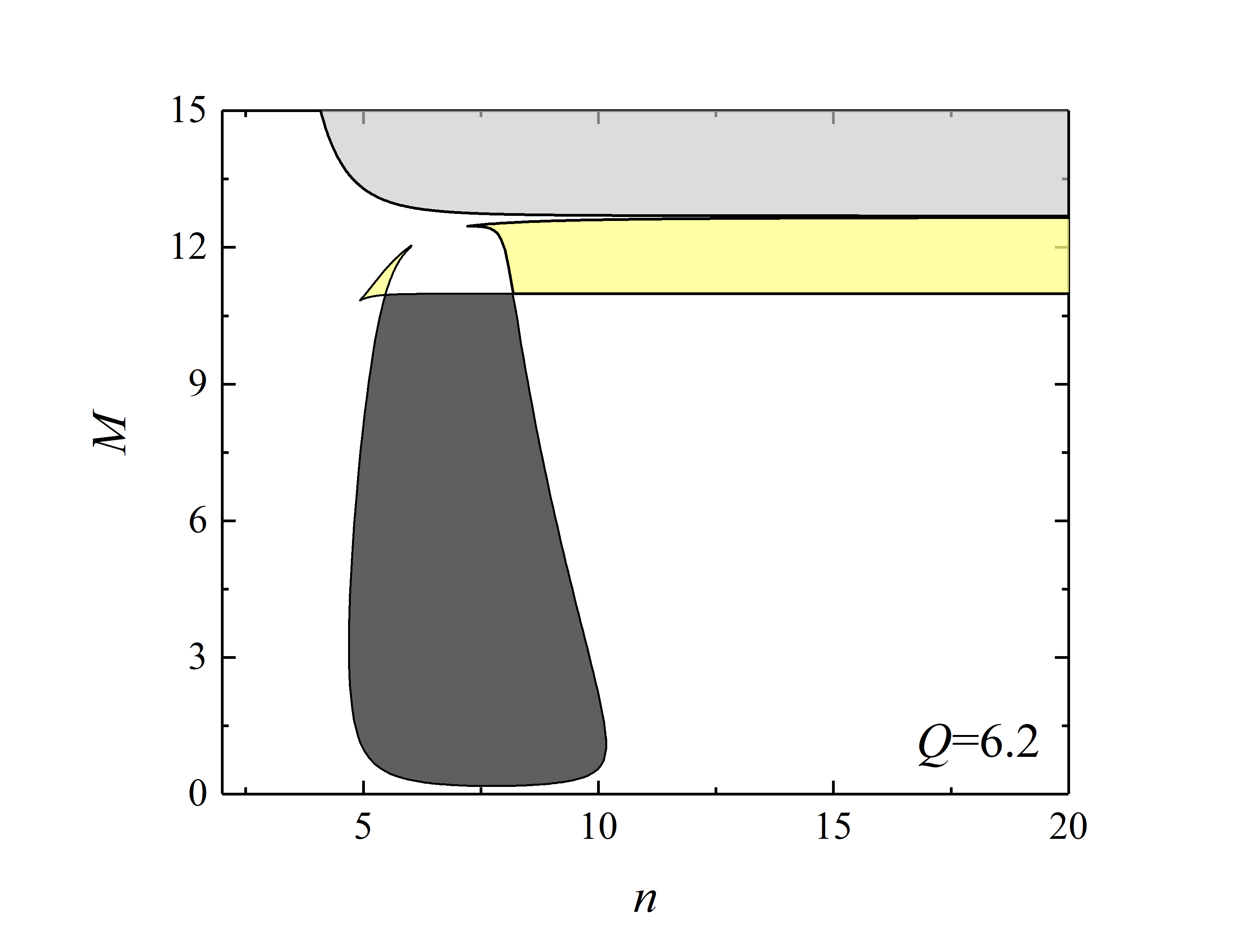}
    \caption{Domains of parameters on the $(n,M)$ plane for different values of $Q$. }
    \label{fig:nM1}
\end{figure}
\begin{figure}
\includegraphics[width=70mm]{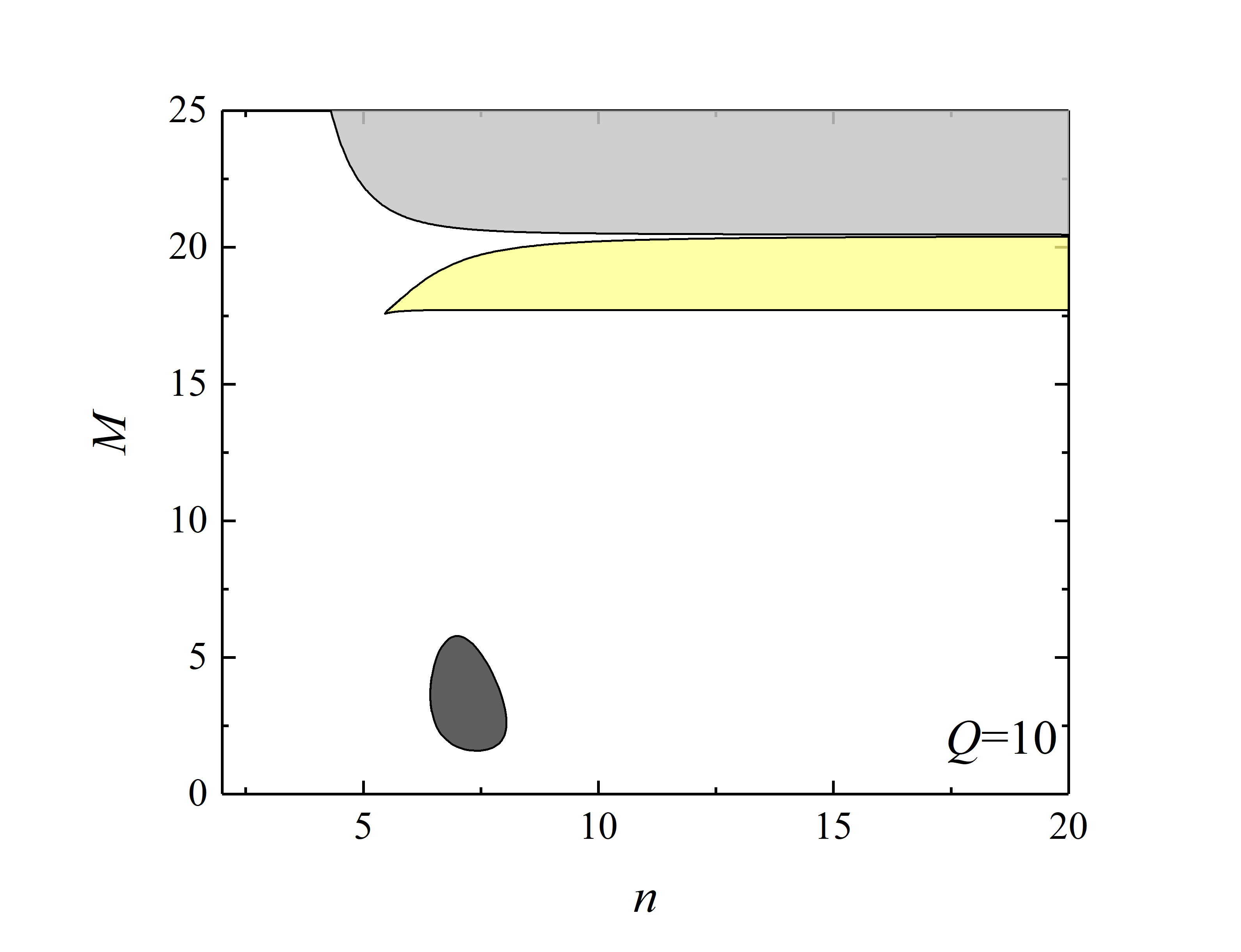}
\includegraphics[width=70mm]{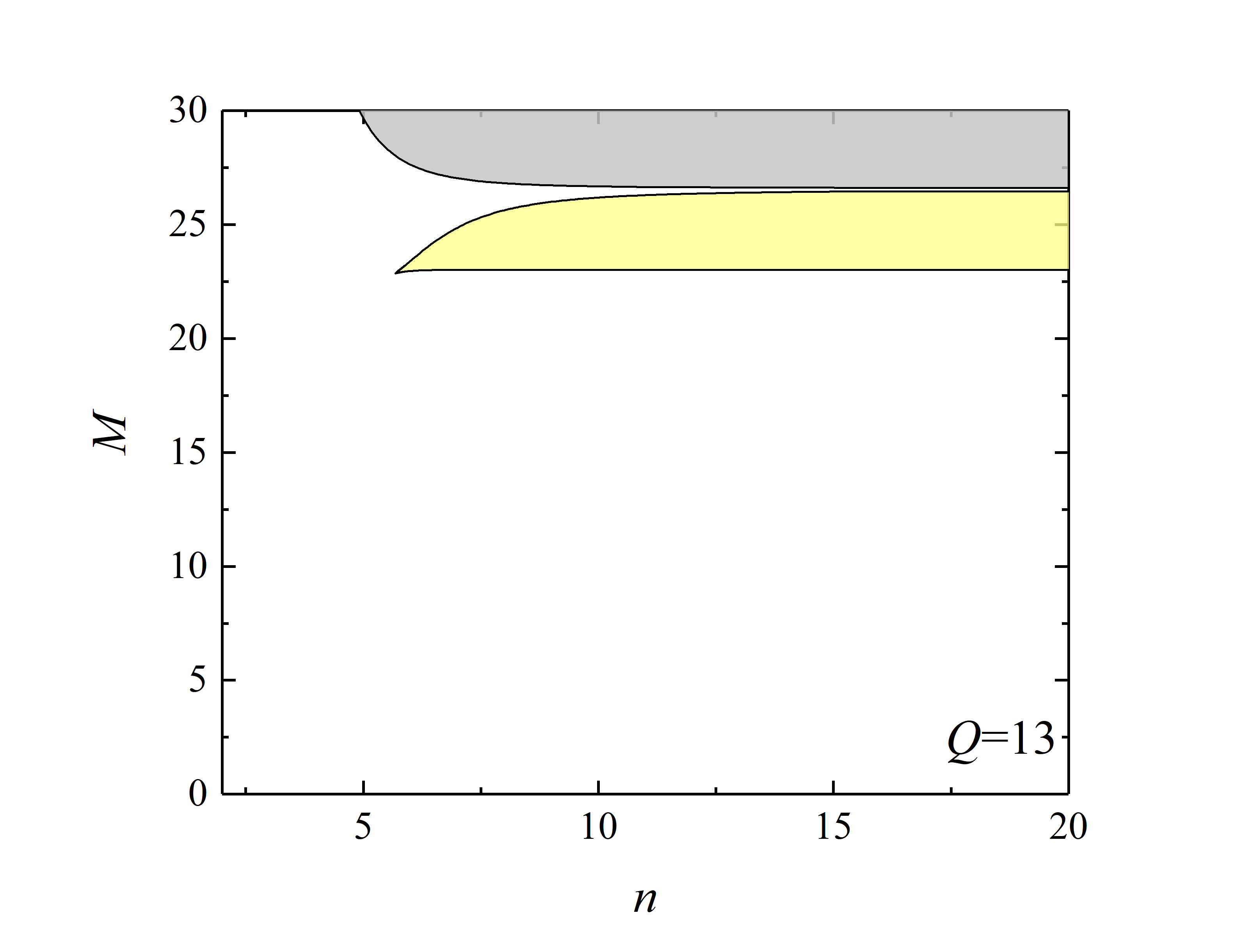}
    \caption{The same as on the previous figure for larger $Q$.  
    }\label{fig:nM2}
\end{figure}

\begin{figure}
\includegraphics[width=70mm]{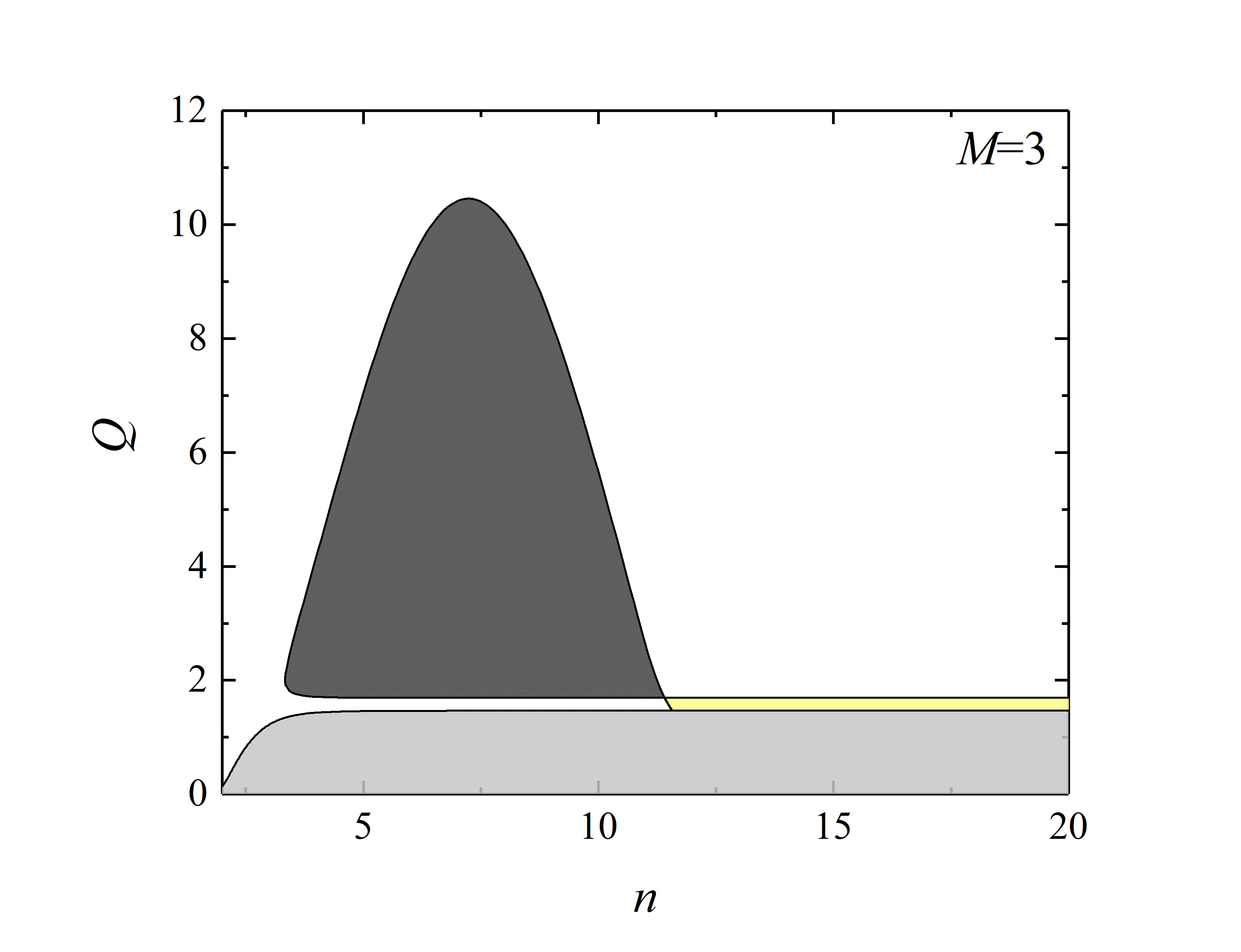}
\includegraphics[width=70mm]{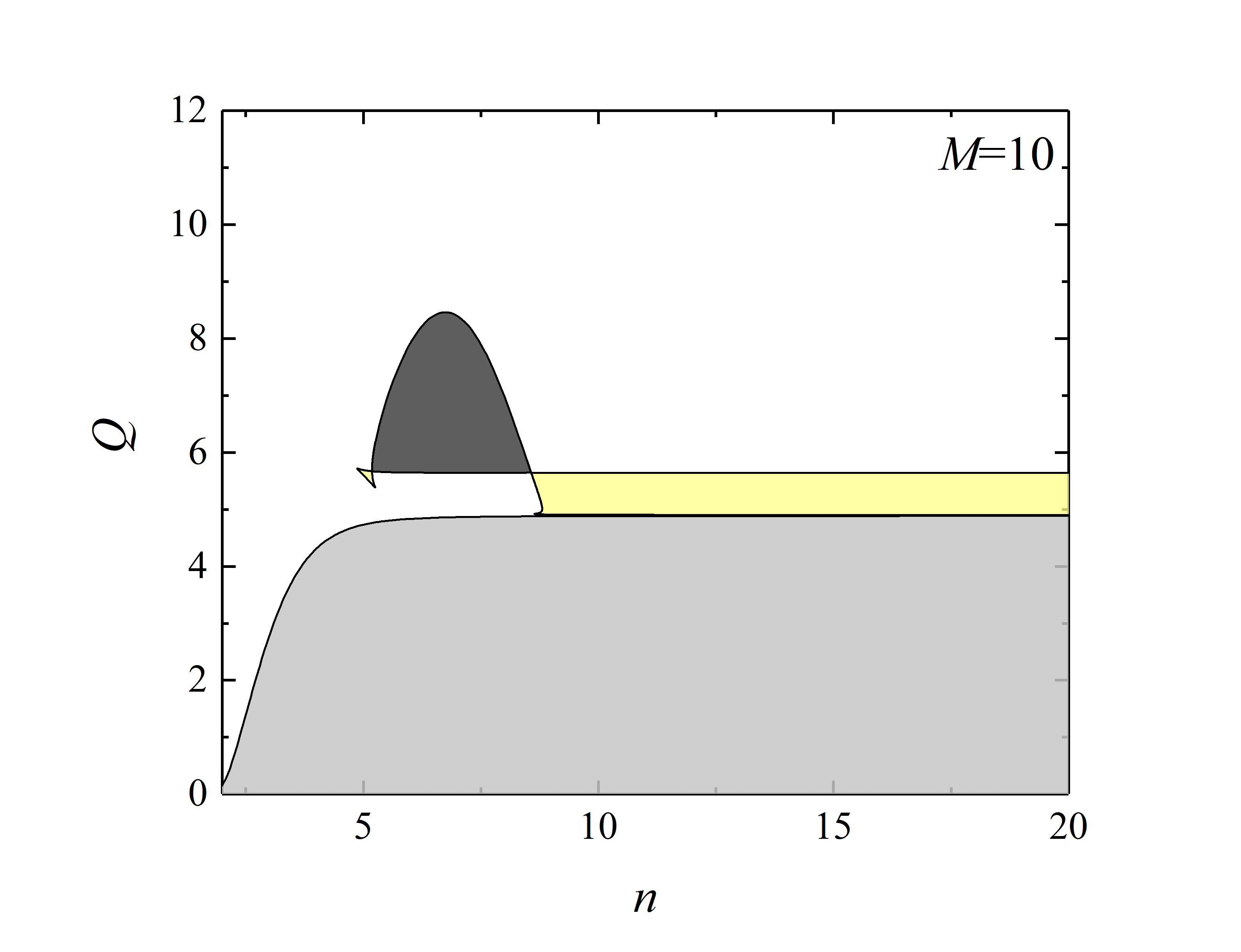}
   \caption{Domains of parameters on $(n,Q)$ plane for different values of $M$.   }
    \label{fig:nQ1}
\end{figure}
\begin{figure}
\includegraphics[width=70mm]{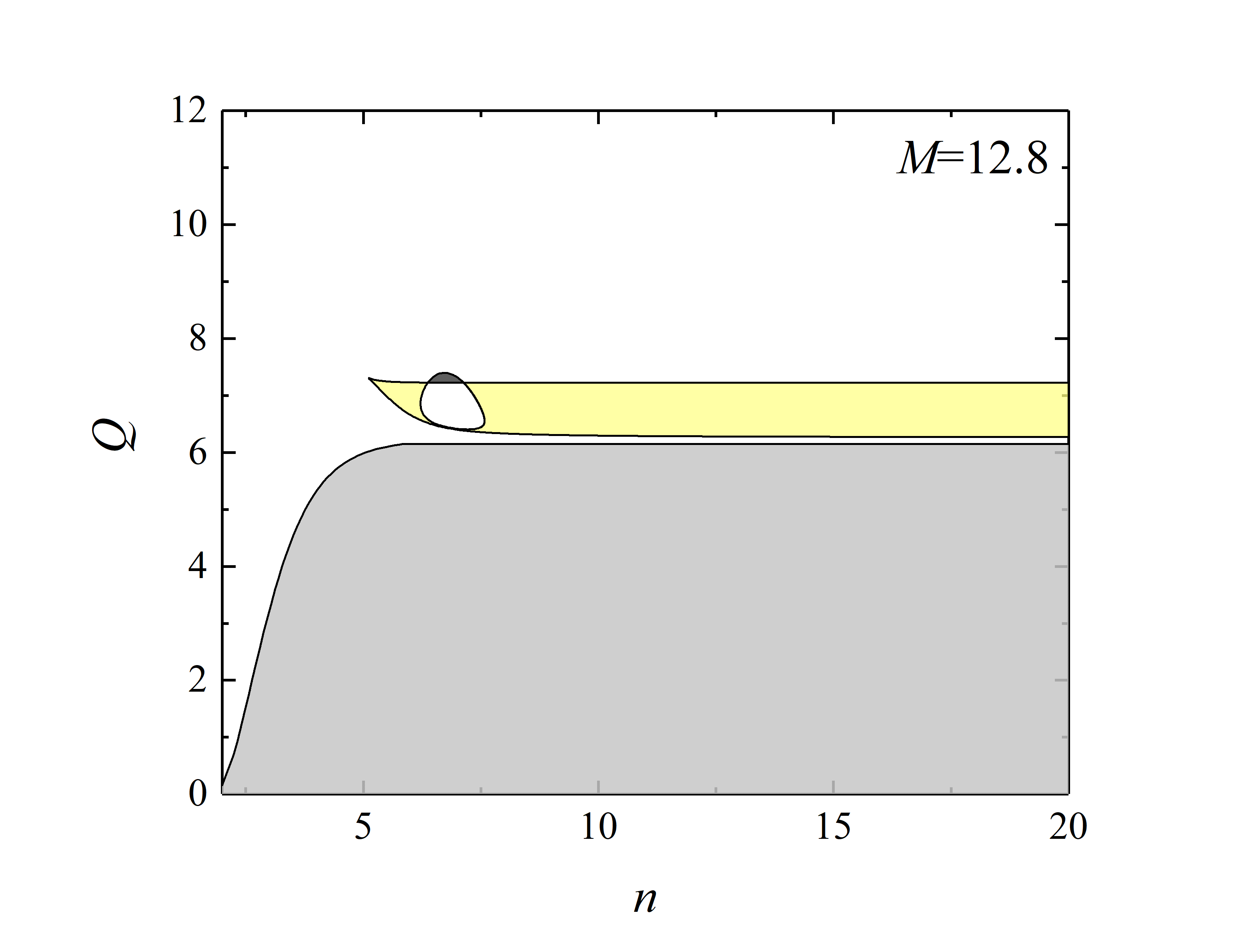}
\includegraphics[width=70mm]{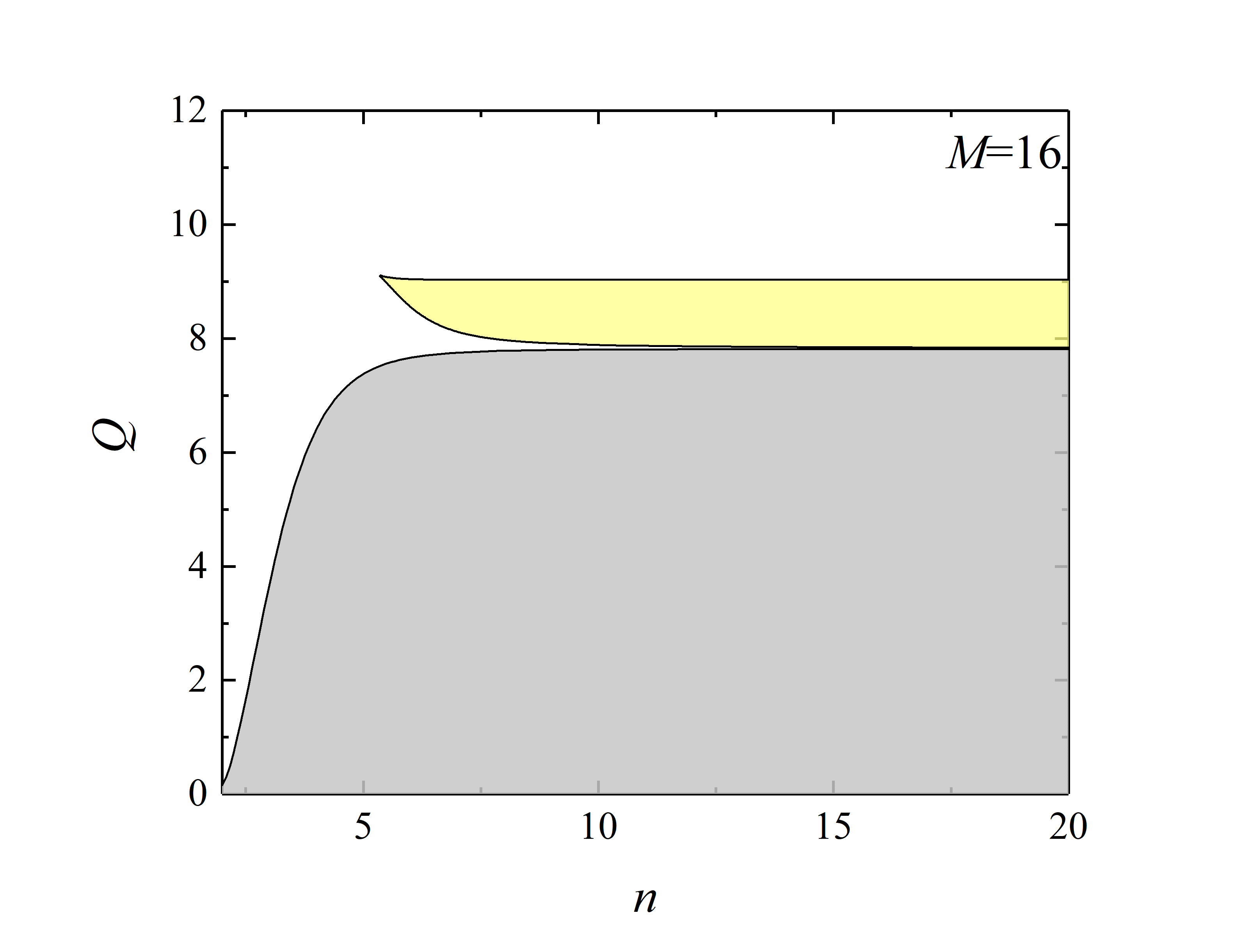}
    \caption{The same as on the previous figure  for larger $M$. For arbitrary $M$ there is a   domain on $(n,Q)$-plane where $S3$ type exist. }
    \label{fig:nQ2}
\end{figure}


\subsection{Photon trajectories and accretion disk images}\label{sub_photon trajectories}
\begin{figure}[h!]
    \centering
\includegraphics[width=70mm]{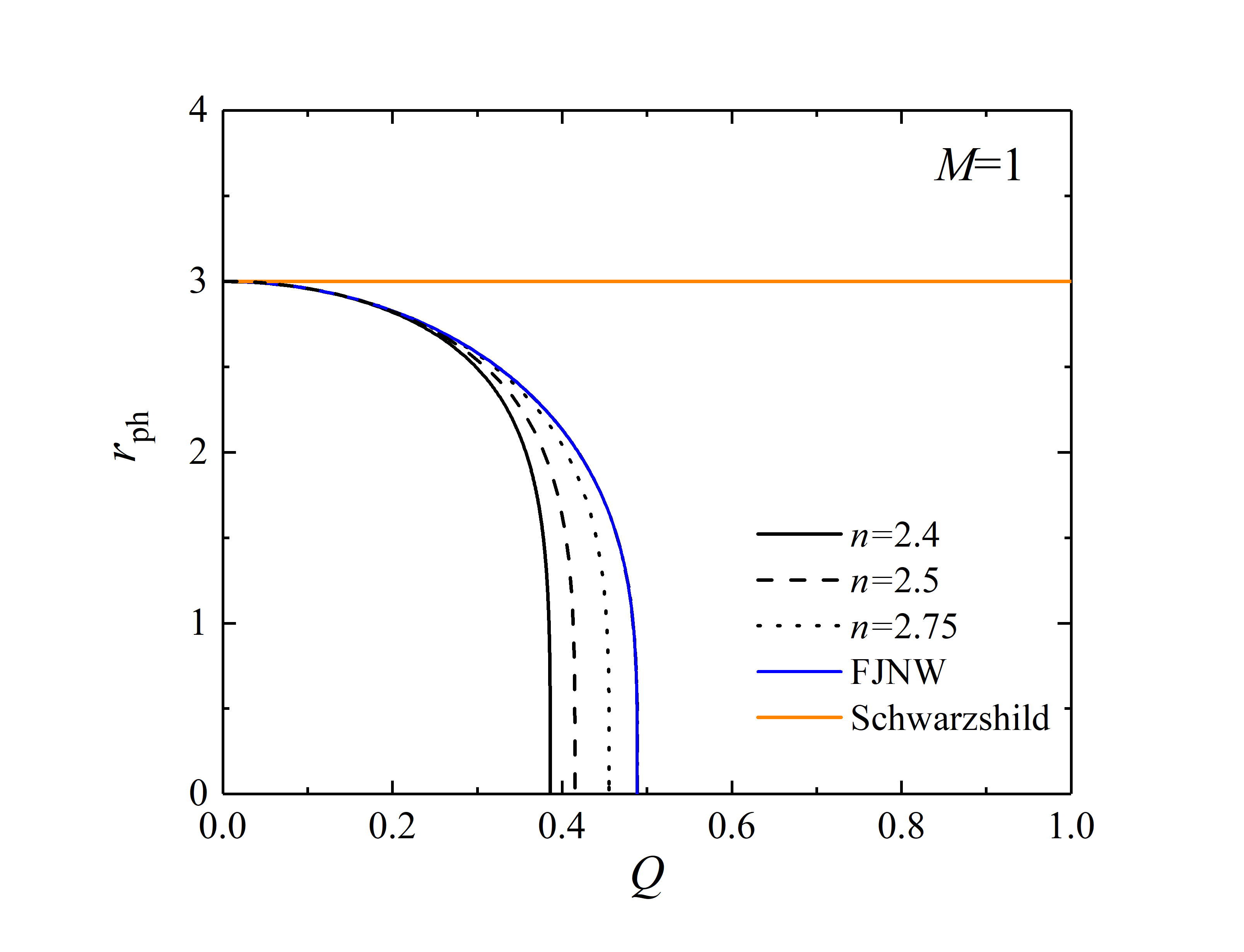}
\includegraphics[width=70mm]{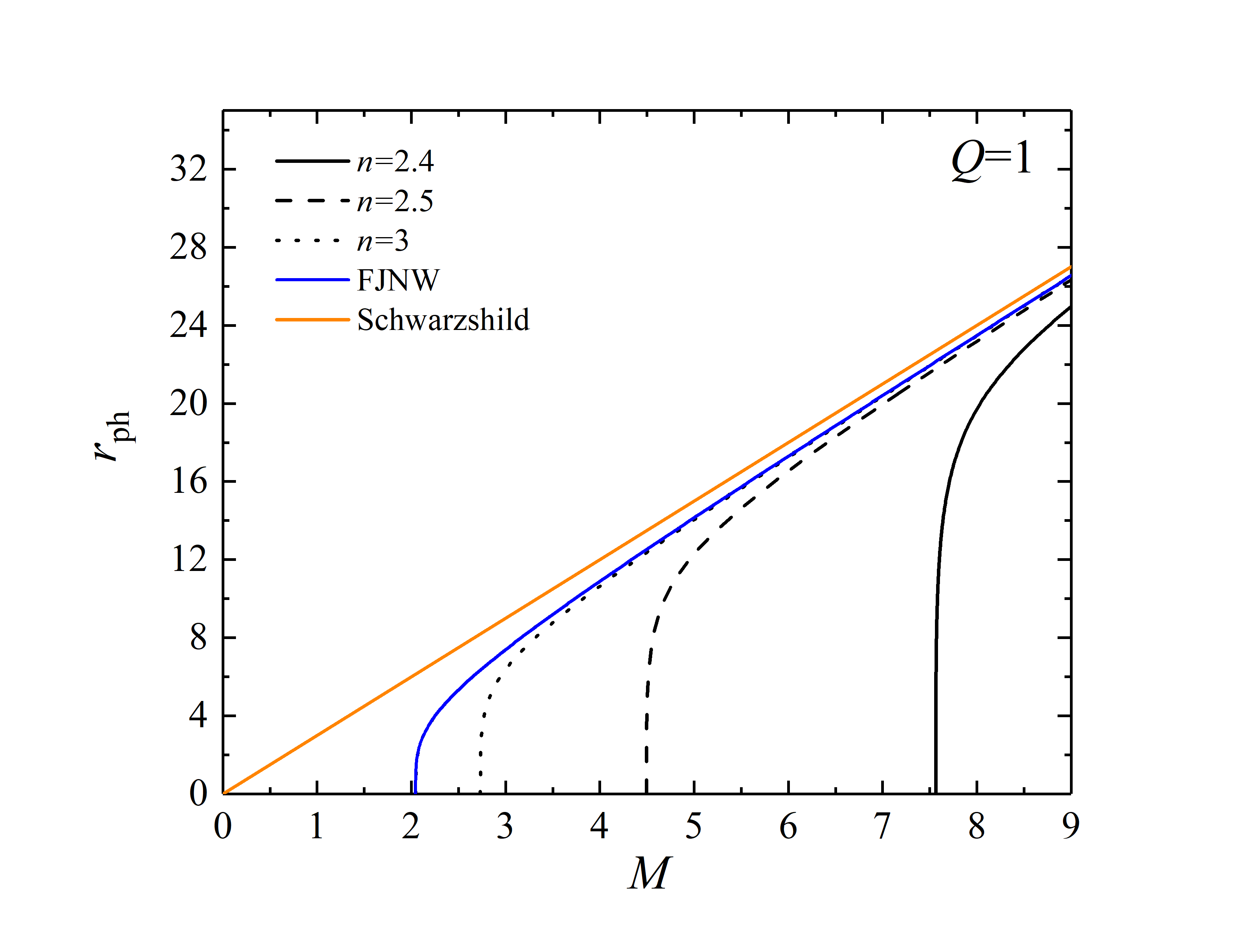}
    \caption{The  photon orbits radii as a function of scalar "charge" $Q$ (left) and configuration mass $M$ (right) for different values of $n$. Blue and orange curves correspond to the cases of FJNW    and Schwarzschild metrics. Curves $r_{\rm ph}$  go below  FJNW  and  tend to them with increasing $n$.}
    \label{fig:PhO-radius}
\end{figure}

In case of null geodesics ($S=0$), the effective potential $U_{\rm eff}(r,L,0)=e^{\alpha}{L^2}/{r^2}$ has simple properties. Its asymptotics  are due to (\ref{U_eff-asymptotics}).  The sign of $U_{\rm eff}/dr$ is defined by function
\be
f(r)= r\alpha'(r)-2
\ee
We have verified numerically  that $f(r)$ is a monotonically decreasing function; evidently $f(\infty)=-2$. For $r\rightarrow{0}$ we have  $f(r)\gtrless0$ if $\eta\gtrless3$. Therefore, the root of this function  can exist only if  $\eta>3$; then the point of maximum $r_{\rm ph}$ is a single root of $f(r)$, $r_{\rm ph}$ being the radius of the photon sphere. 
Fig. \ref{fig:PhO-radius}  shows typical dependencies of the photon orbit`s radii on the configuration parameters. One can see that $r_{\rm ph}$ is always  less than the  corresponding radius in Schwarzshild/FJNW cases.

The next step in the study of the configurations in question concerns the images of accretion discs represented by different types of SCOD. 
 In order to build direct accretion disk images, we use the ray-tracing algorithm described in \cite{Psaltis_2011,JohannsenJ,Bambi_2012}. 
 The complete consideration of this problem requires knowledge of the brightness distribution of radiating matter over the disk. We have estimated the surface brightness within the Page-Thorne model \cite{Page_Thorne} showing the there is a great enhancement of the radiation flux from the innermost region with  small radii normalized to the mass accretion rate $\dot M_0$ (analogously to the results of \cite{Chowdhury_2012,Gyulchev_2019} on FJNW solution). However, the (unknown)  value of $\dot M_0$ may be very different   for outer and inner  SCO rings and one can expect that the input of the inner ring will be much smaller due to a scattering of the accreting material. This requires a significant modification of the AD model, presumably within the framework of full-scale hydrodynamic modeling, which is beyond the scope of this article. In this view, we limited  ourselves  to show the  observed  contours of the SCO regions for accretion disks and the frequency distributions over these disks due to the gravitational redshift and the Doppler effects.
 
\begin{figure}
    \centering
\includegraphics[width=80mm]{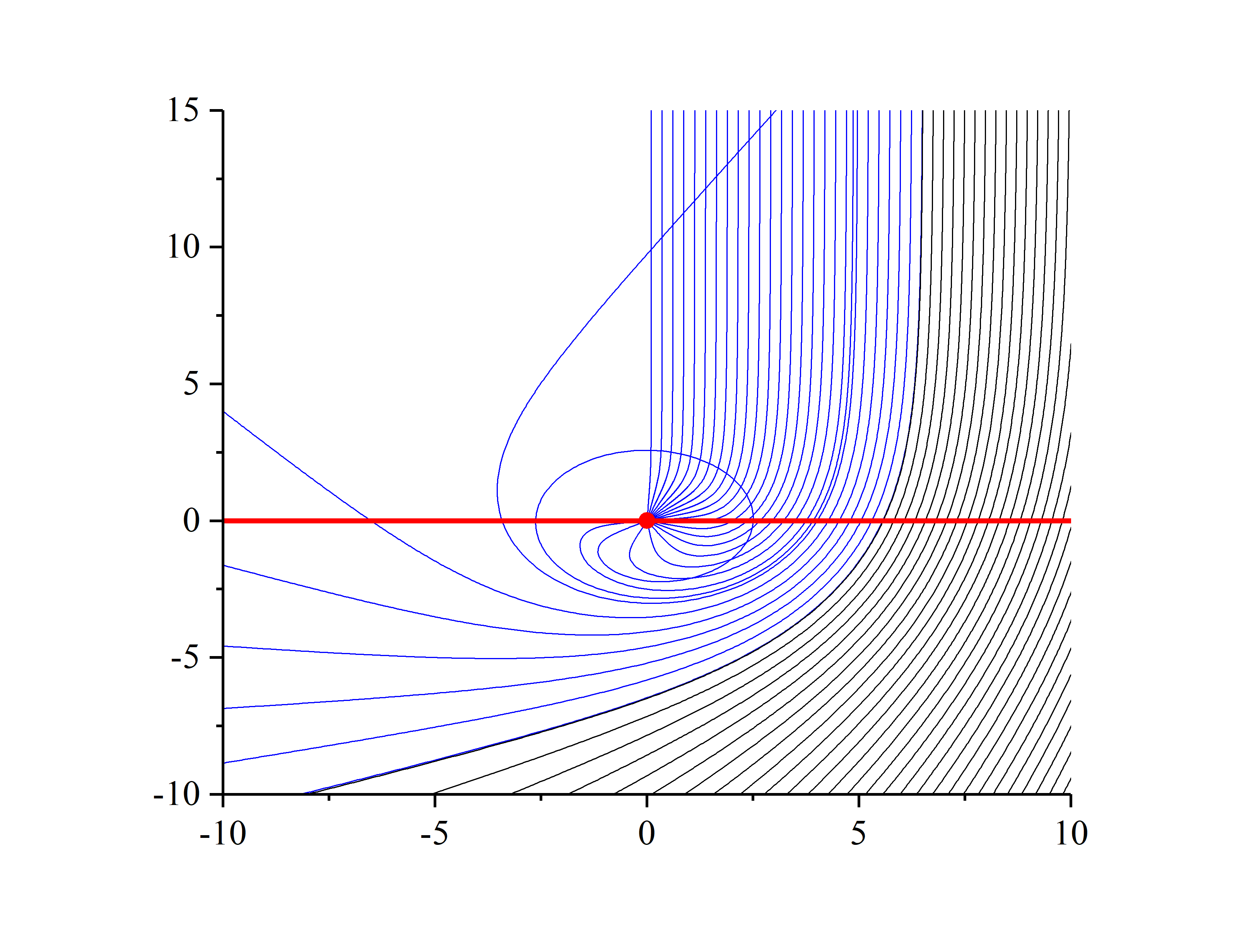}
\includegraphics[width=80mm]{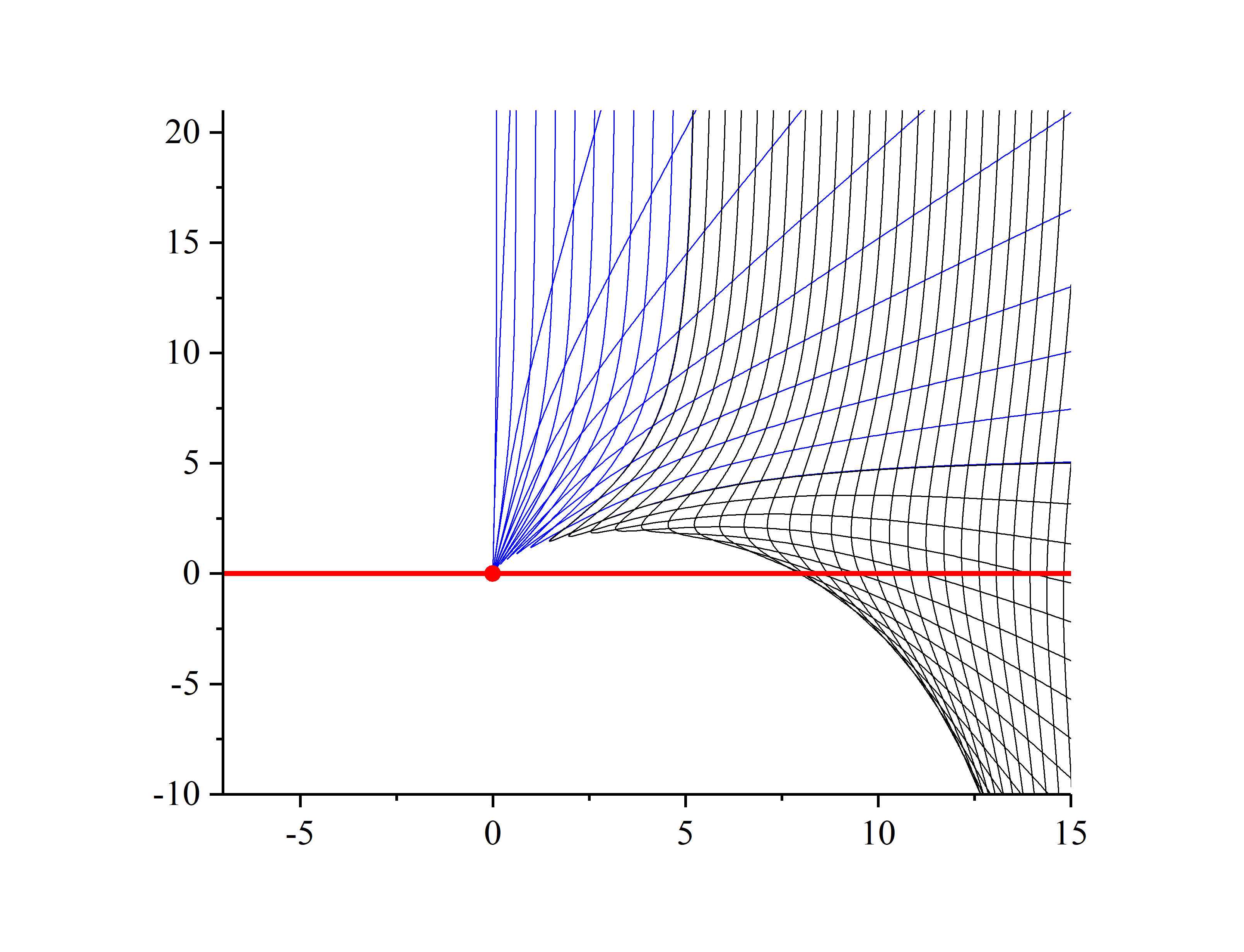}
    \caption{In both panels $n=3$,   $M=1$, the red line indicates the equatorial plane. Left: trajectories of  photons incident from infinity to the attracting   singularity; $Q=0.3$, $\eta=4.89$. Right: the same with  the repulsing singularity $Q=2.2$,   $\eta=2.33$; in this case, the points near the singularity  form a dark spot around the center, imitating a black hole.   }
    \label{photons}
\end{figure}

To plot the  SCOD images as seen by a distant observer, we need the trajectories of photons falling from infinity.
Their properties depend on the sign of $\eta-3$. For $\eta>3$, when there is  maximum of $U_{\rm eff}$, then the incoming photons with impact parameter $\lambda<[b_{max}]^{-1/2}$, where $b_{max}=\exp{[\alpha(r_{\rm ph})]}/r^2_{\rm ph}$, will reach singularity at the origin (see the left panel of Fig. \ref{photons}).  For $\eta<3$ the photons with a nonzero angular momentum will be  reflected  from the  potential (right panel of Fig. \ref{photons}). Due to the strong bending of the rays, a scattered photon can hit a point on the AD plane far enough from the center, where another photon with a different trajectory also hits.  Each such point has two images. On the other hand, there are no photons falling into the AD region  near the singularity; this area is not visible to a distant observer.

 The photon trajectories needed in the ray-tracing method were obtained numerically, similar to those shown in Fig. \ref{photons}.
We fix a sufficiently large distance to the static observer, where the geometry can be considered flat, and we track the photons coming from the observer to the AD plane,  where we take into account only those photons that hit the SCO regions. We do not take into account the input from the reverse side of AD.  The frequency ratio $g$ between the point $(e)$ at the AD surface and static remote observer $(o)$ for metric \eqref{metric} is 
\begin{equation}
g=\frac{k_{\mu}u^{\mu}|_o}{k_{\mu}u^{\mu}|_e}=\frac{\sqrt{e^{\alpha(r)}-r^2\Omega^2}}{1+\lambda\Omega}.
\end{equation}
This formula takes into account both gravitational and Doppler effects. We use the  normalized redshift factor
\begin{equation}
\tilde {g}=\frac{g-g_{\rm min}}{g_{\rm max}-g_{\rm min}},
\end{equation}
where $g_{\rm min}$ and $g_{\rm max}$ is the minimal and maximal frequency ratios on disk, respectively. 

\begin{figure}
    \centering
\includegraphics[width=80mm]{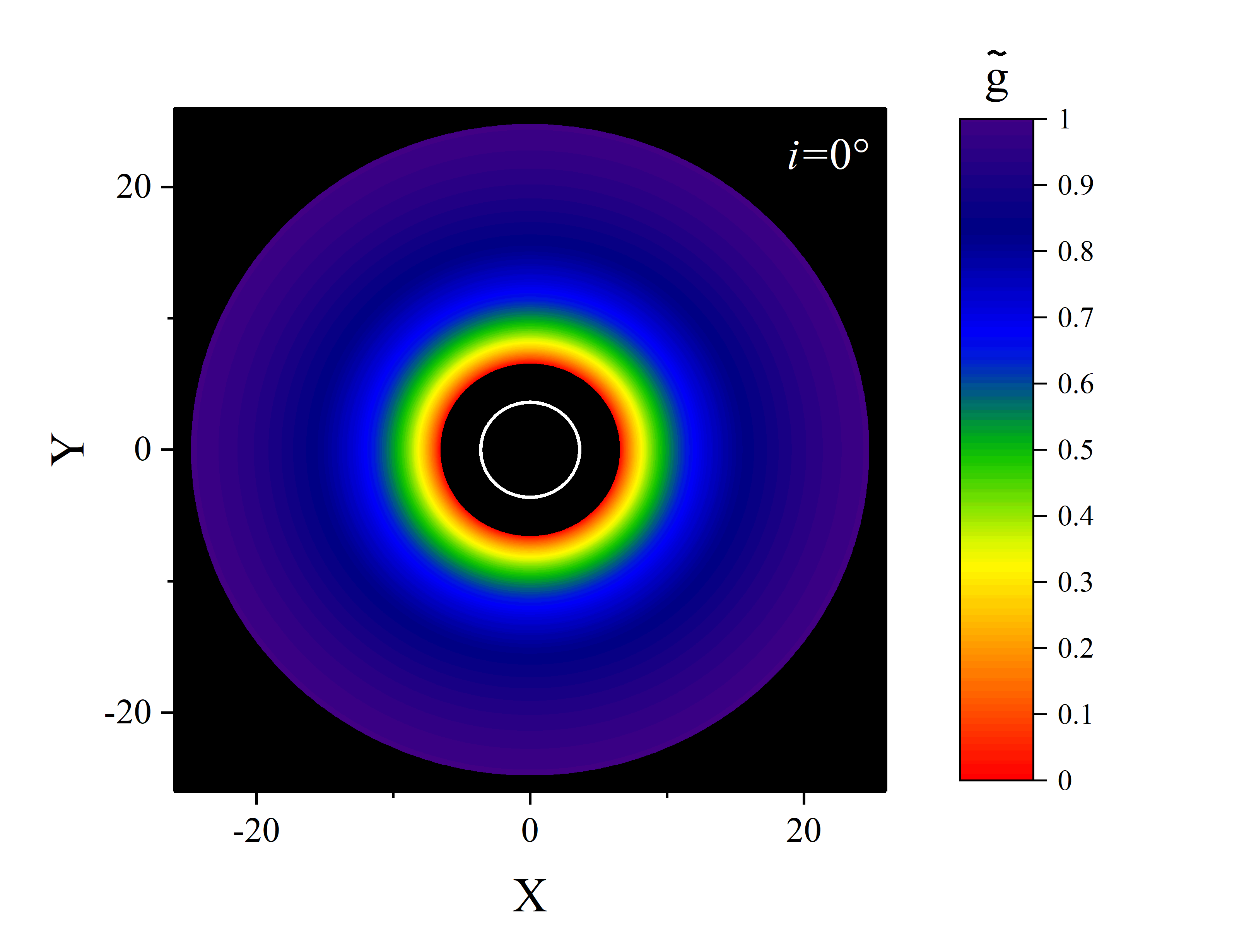}
\includegraphics[width=80mm]{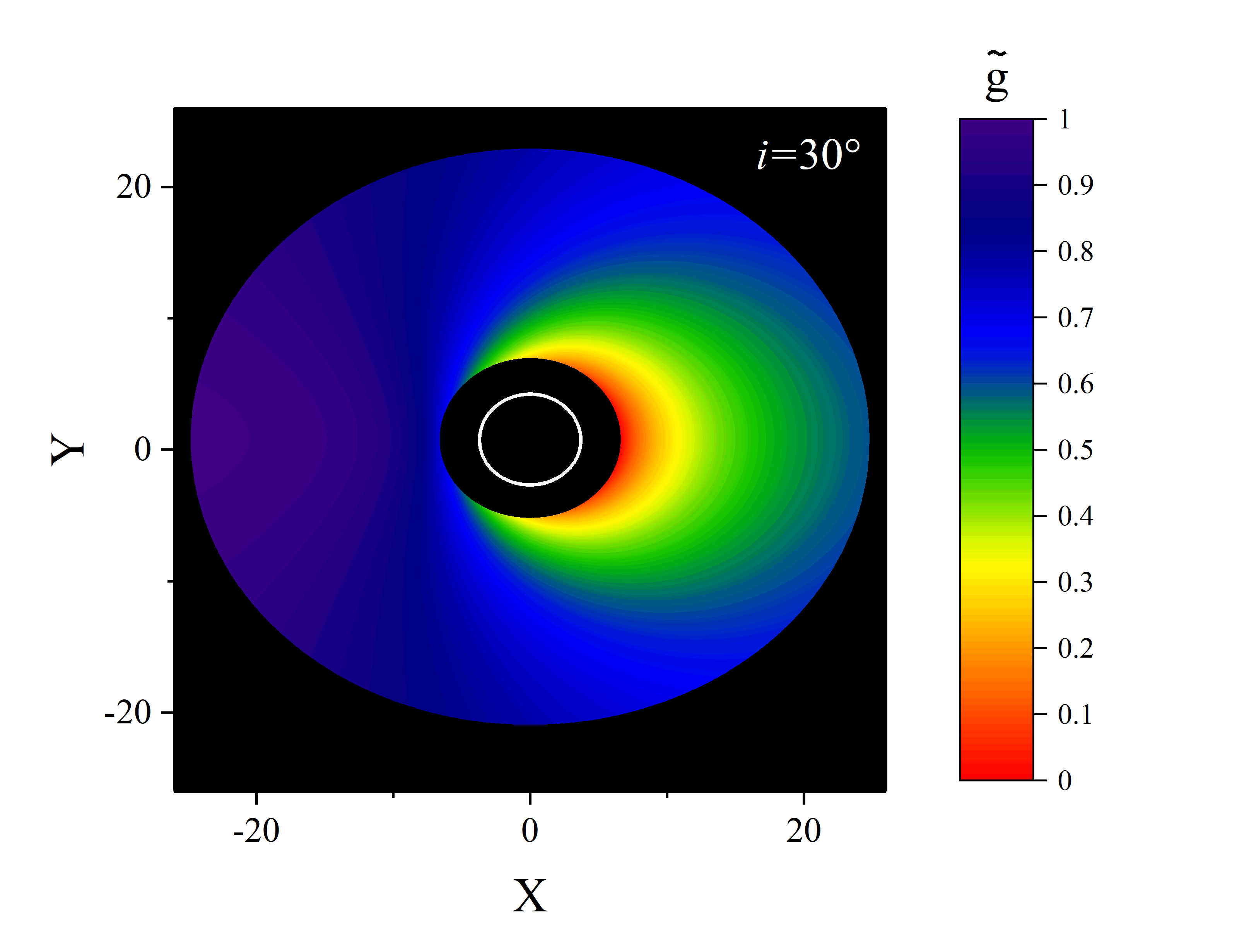}
    \caption{AD images for the $U1$ type of SCOD ($M=1$, $Q=0.3$, $n=3$): in full face and  for   inclination $i=30^o$.  White contour corresponds to the photon orbit at $r_{\rm ph}\approx 2.58$. The ISCO placed at $r_{1(U)}\approx 5.56$ and the outer disk edge was fixed at $r=24$. }
    \label{fig:diskU1}
\end{figure}

\begin{figure}
    \centering
\includegraphics[width=80mm]{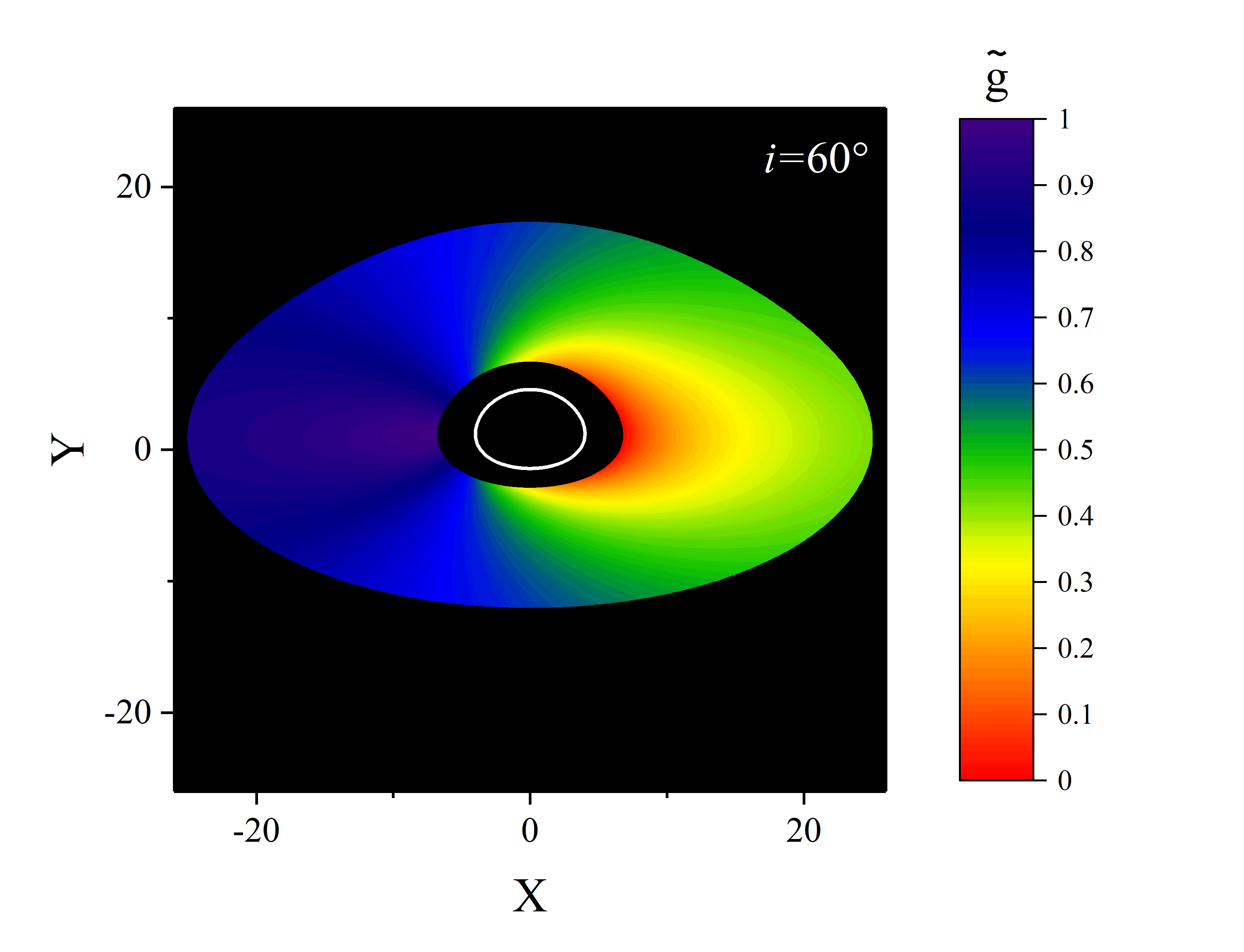}
\includegraphics[width=80mm]{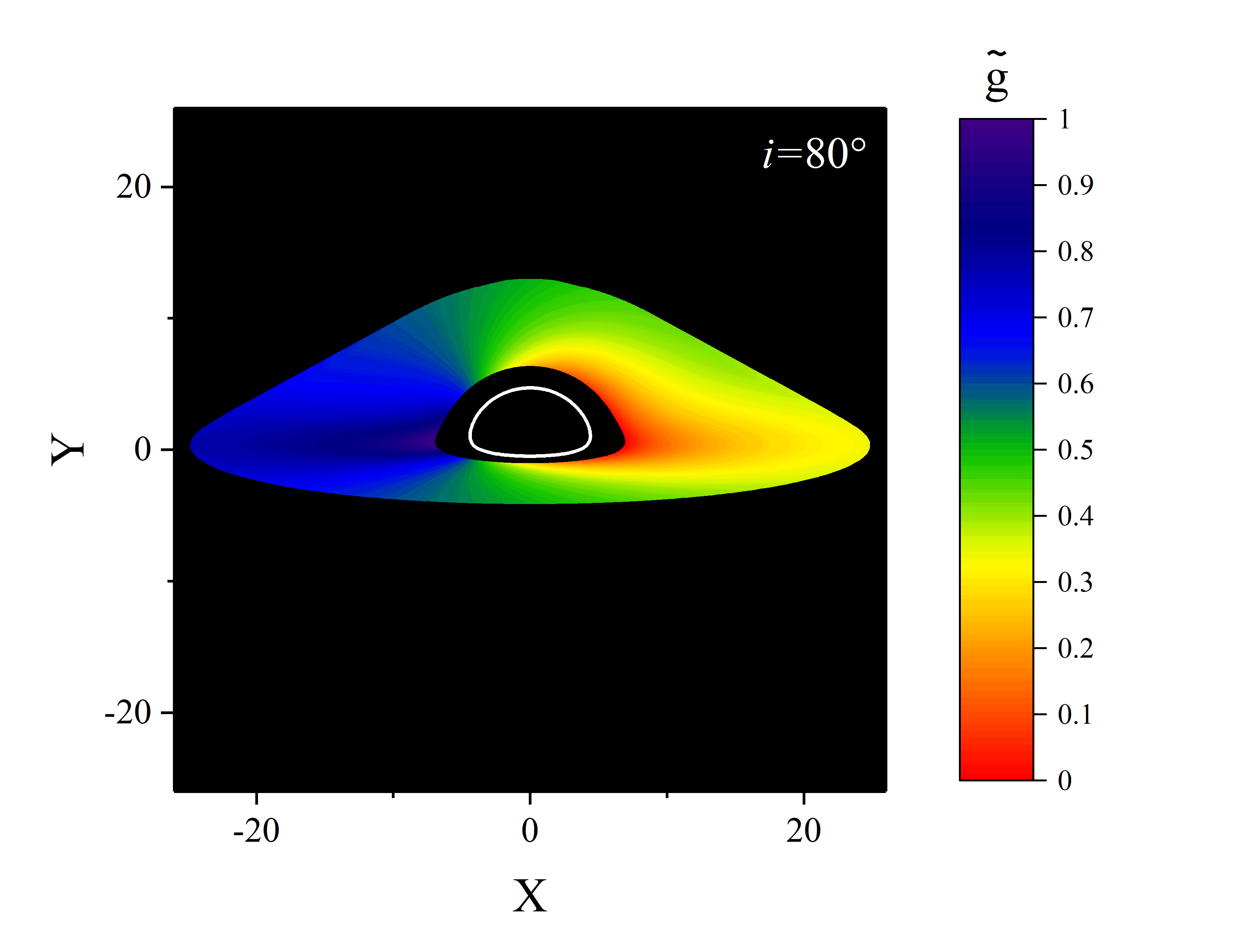}
    \caption{The same as on Fig. \ref{fig:diskU1} with inclinations $30^o$ and $60^o$.   }
    \label{fig:diskU1-b}
\end{figure}

\begin{figure}
    \centering
\includegraphics[width=80mm]{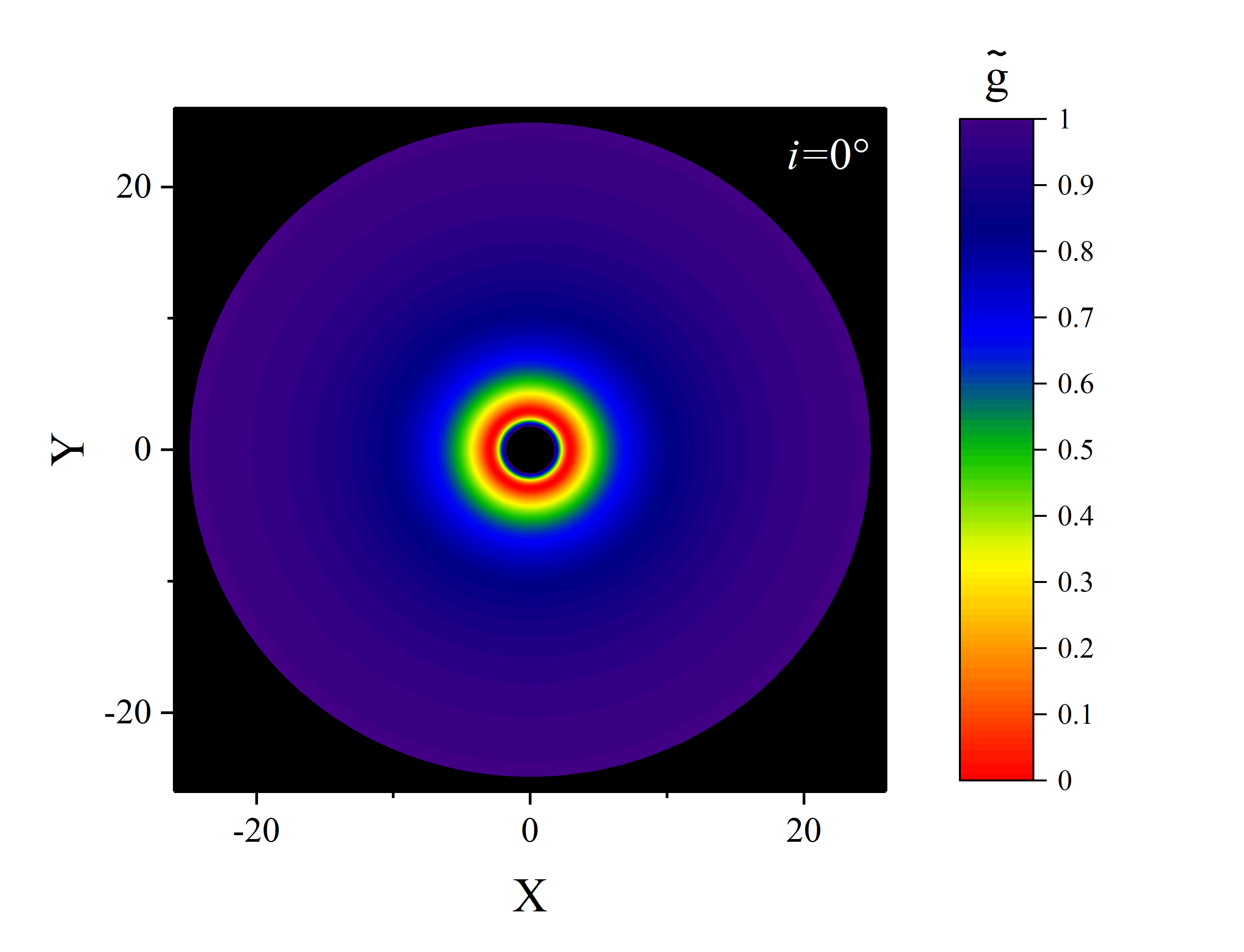}
\includegraphics[width=80mm]{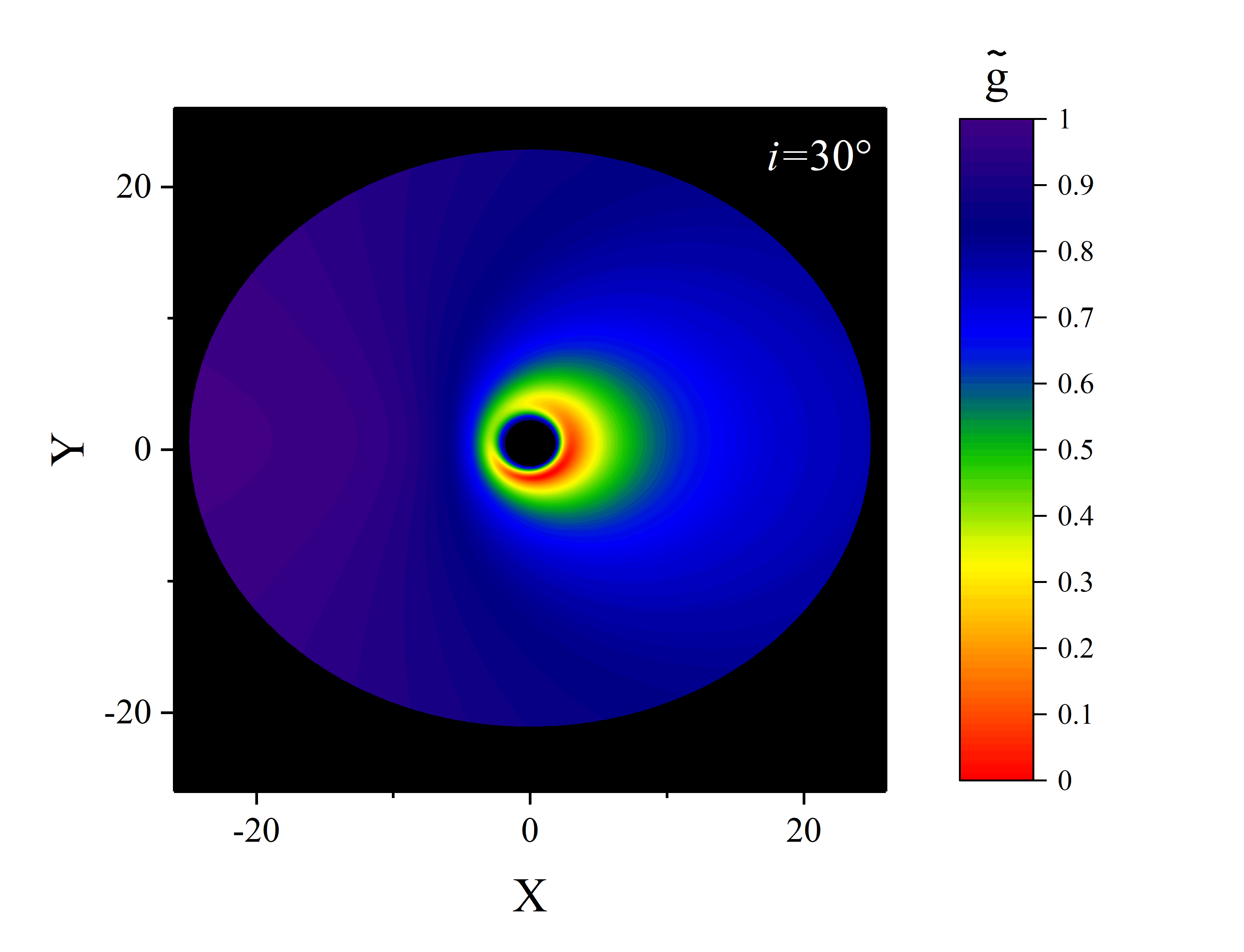}
\caption{AD images for the  $S1$ type ($M=1$, $Q=0.8$, $n=3$); in full face and  for   inclination $i=30^o$.   The outer disk edge radius $r=24$. There is a dark spot in the center due to the repulsive character of the naked singularity.}
    \label{fig:diskS1}
\end{figure}

\begin{figure}
   \centering
\includegraphics[width=80mm]{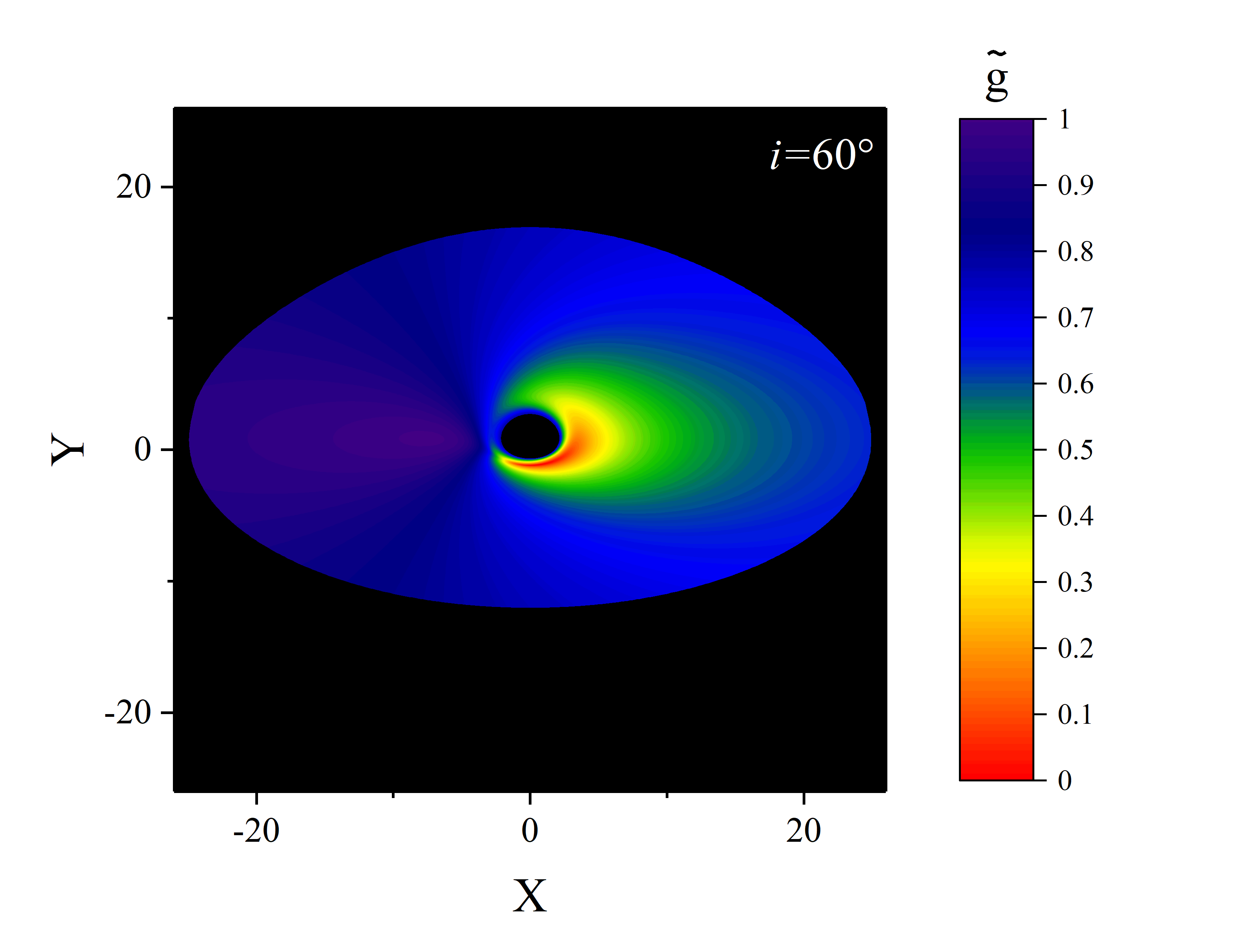}
\includegraphics[width=80mm]{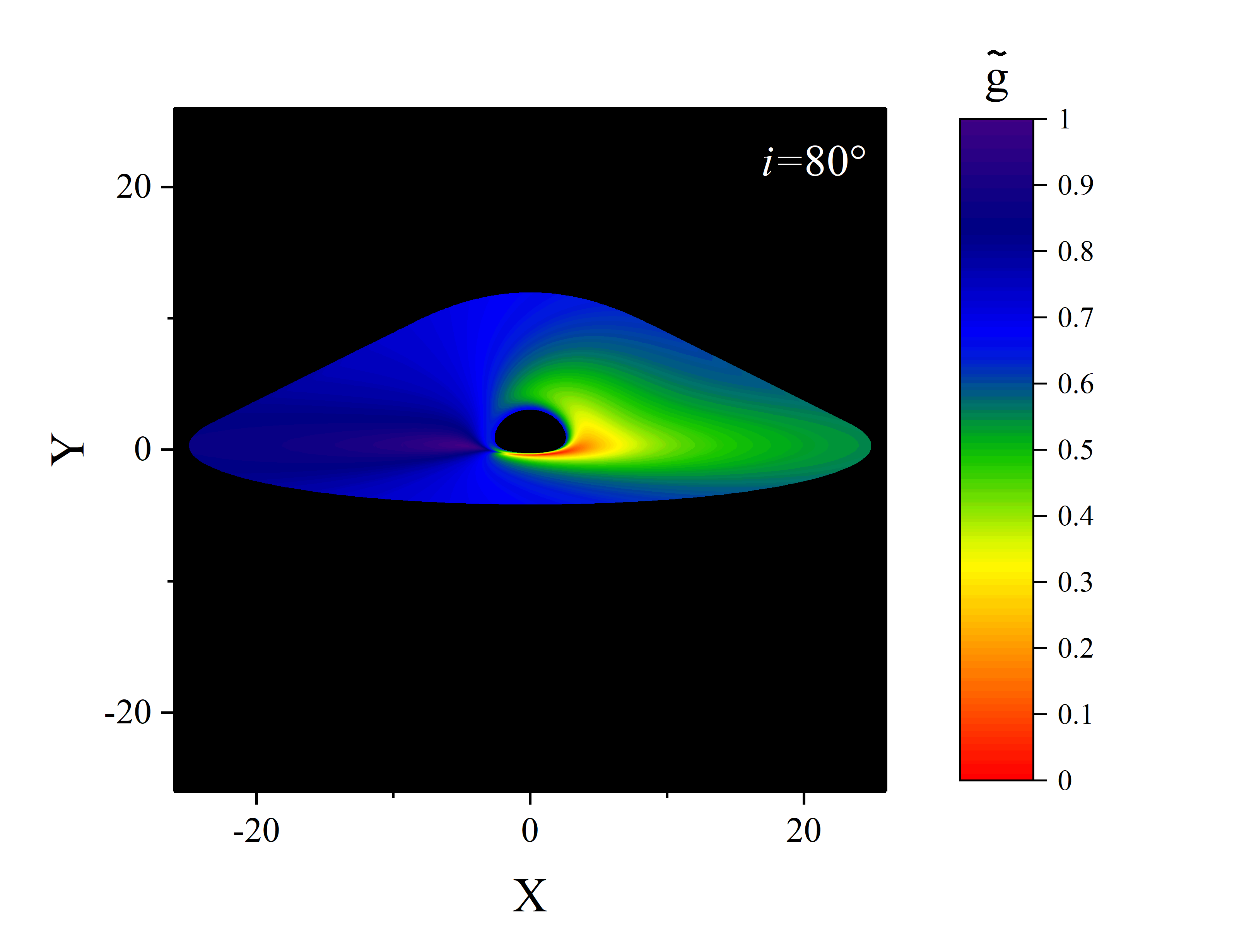}
    \caption{The same as on Fig. \ref{fig:diskS1} with inclinations $60^o$ and $80^o$.   }
    \label{fig:diskS1-b}
\end{figure}

\begin{figure}
    \centering
\includegraphics[width=80mm]{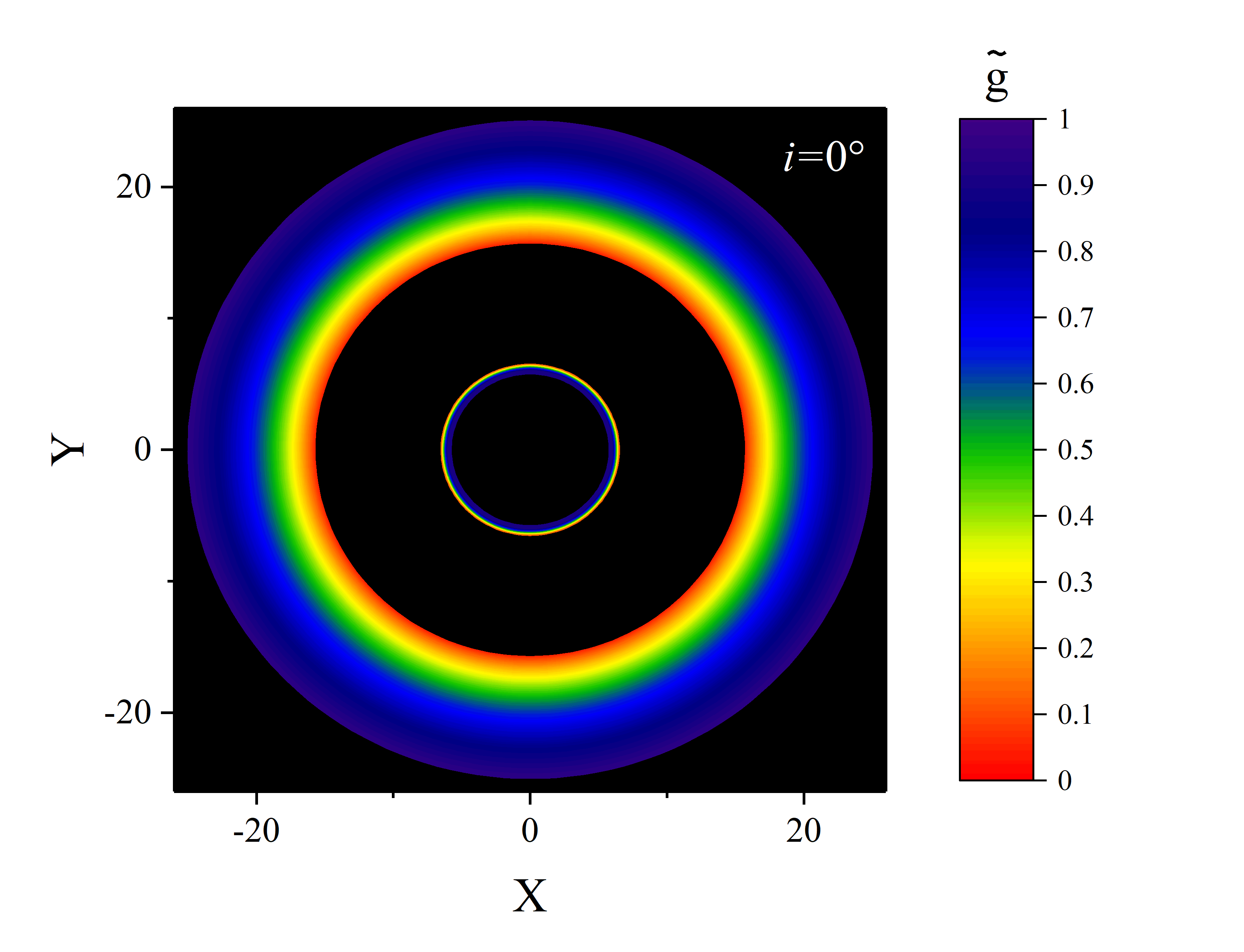}
\includegraphics[width=80mm]{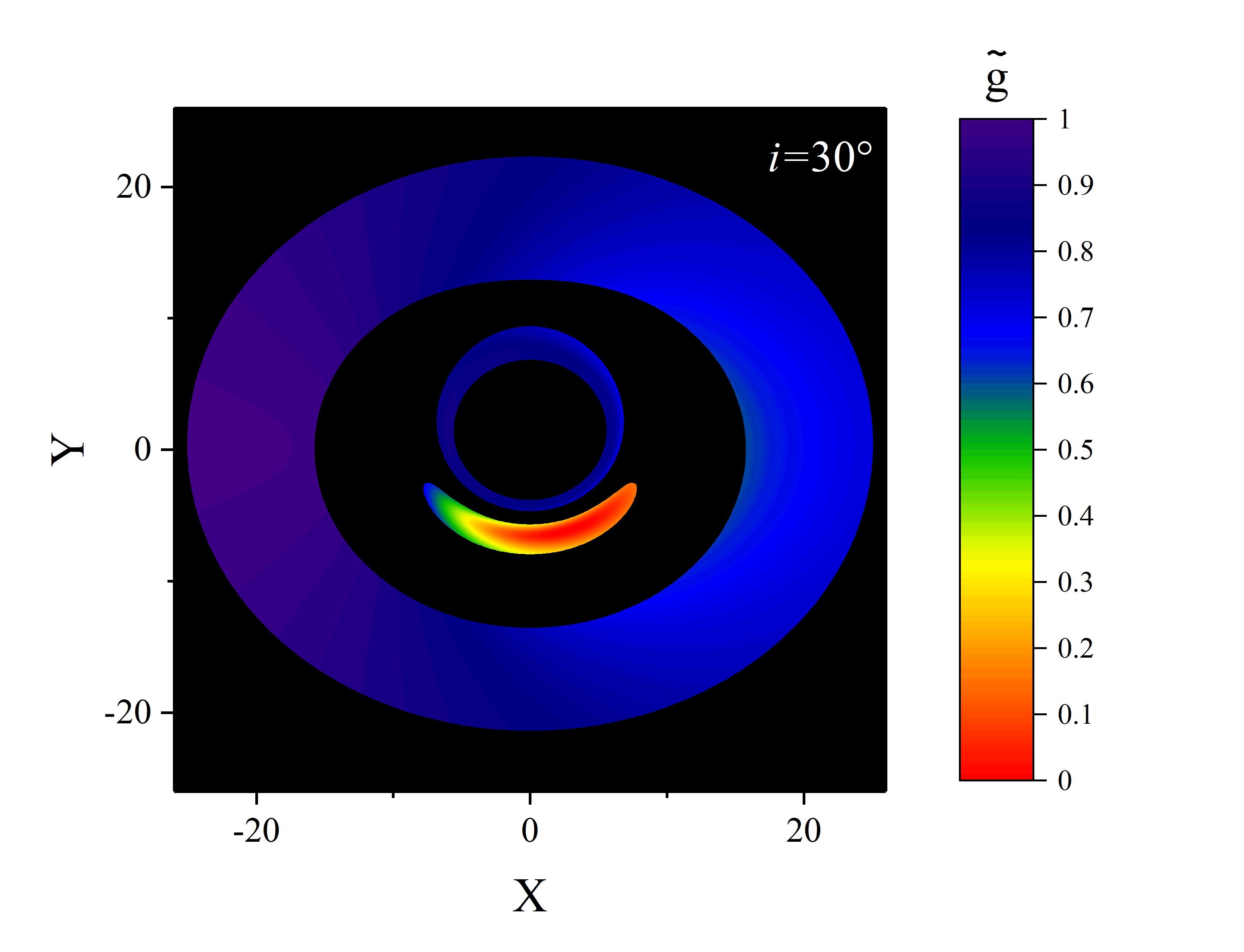}
    \caption{AD images for the  $S2$ type ($M=1$, $Q=2.2$, $n=3$) in full face and  for   inclination $i=30^o$. The outer disk edge radius is $r=24$. SCO radii are in intervals $r\in(0,6.5)$ and $r\in(14.5,\infty)$.   The inner SCO region in the first figure cannot be observed due to the repulsive nature of the naked singularity.}
    \label{fig:diskS2}
    \end{figure}

\begin{figure}
    \centering
\includegraphics[width=80mm]{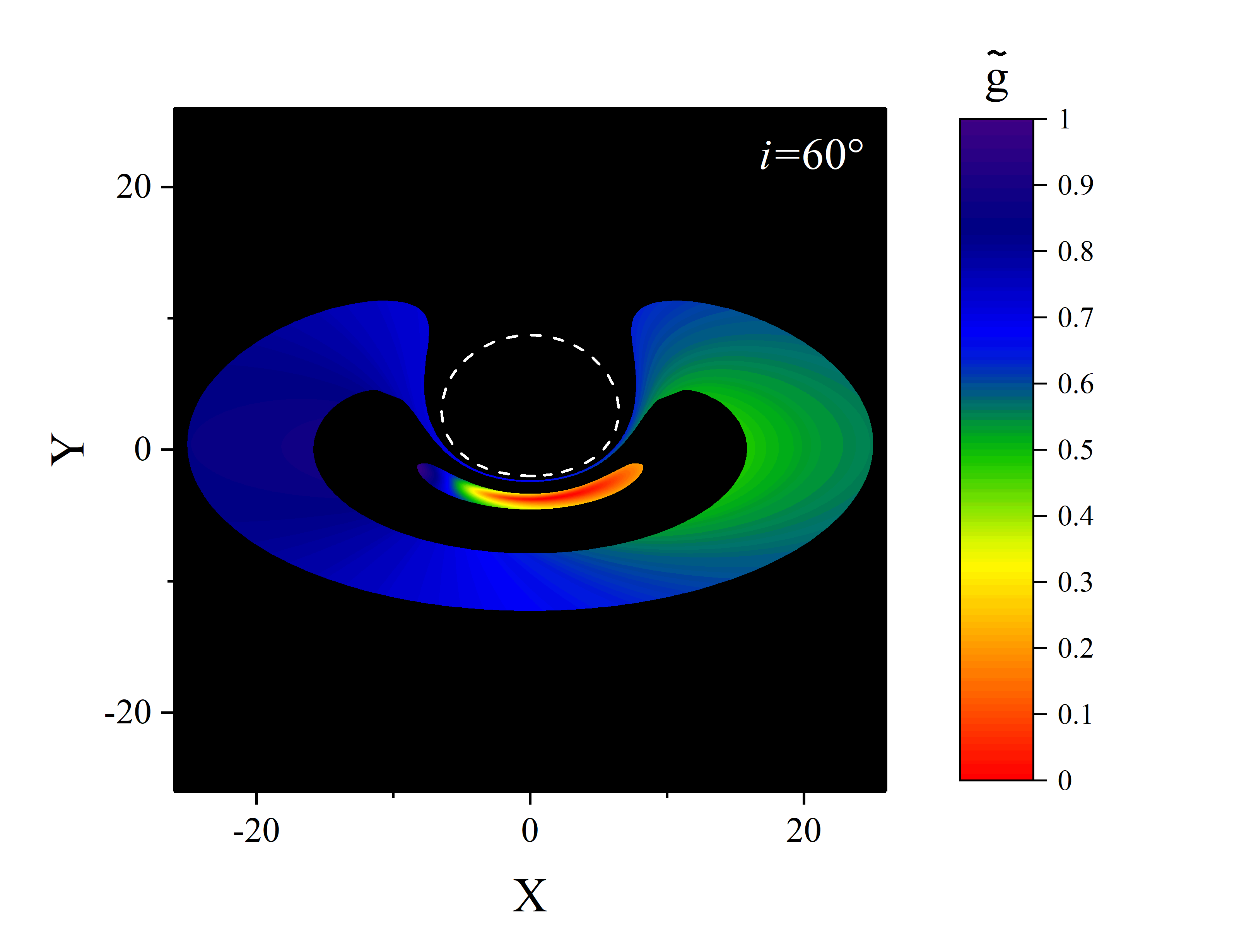}
\includegraphics[width=80mm]{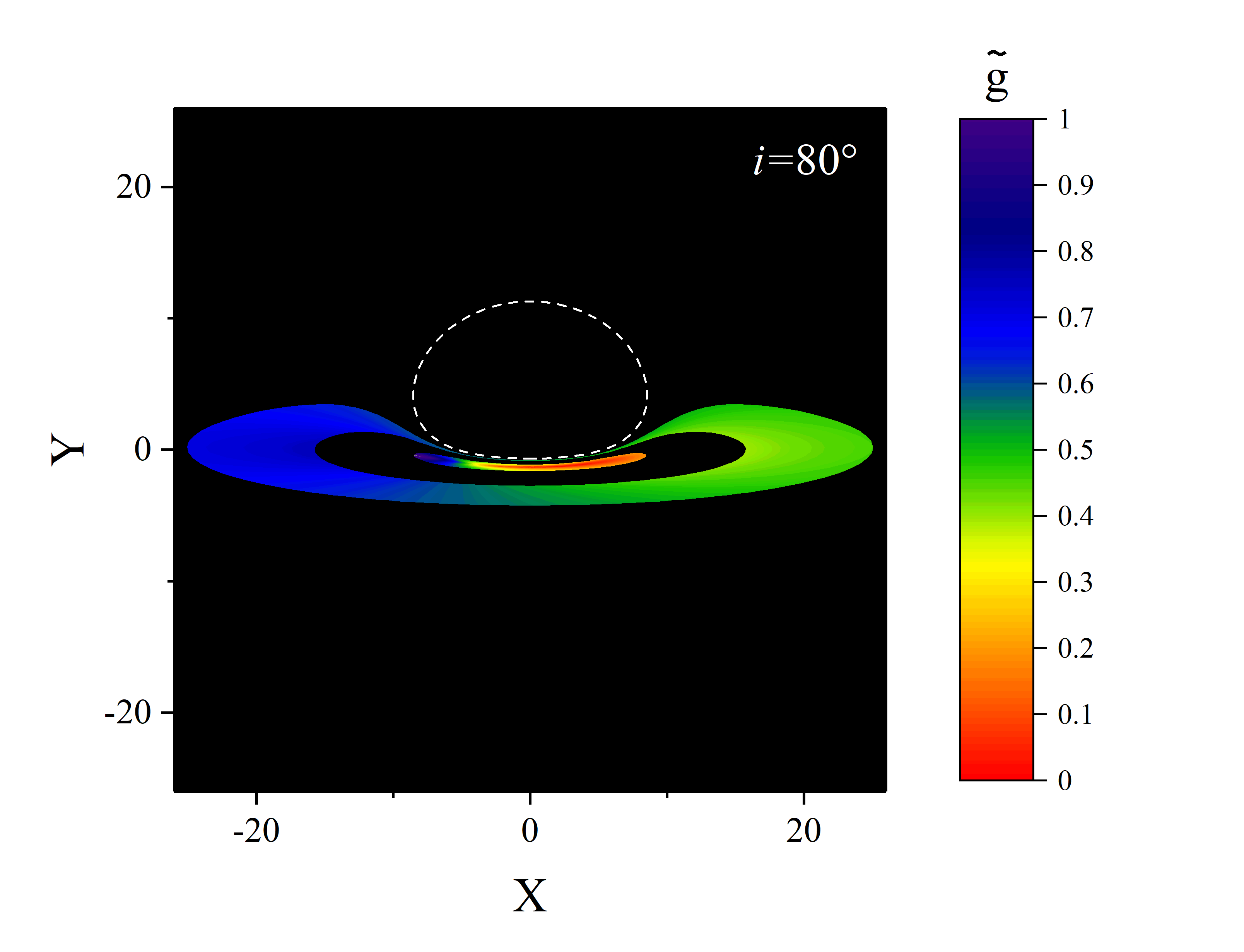}
    \caption{The same as on Fig. \ref{fig:diskS2} with inclinations $60^o$ and $80^o$. The outer disk edge is  $r=24$. White dashed line here shows the outline of the shadow of the naked singularity when it is illuminated by background light.}
    \label{fig:diskS2-b}
    \end{figure}

\begin{figure}
    \centering
\includegraphics[width=80mm]{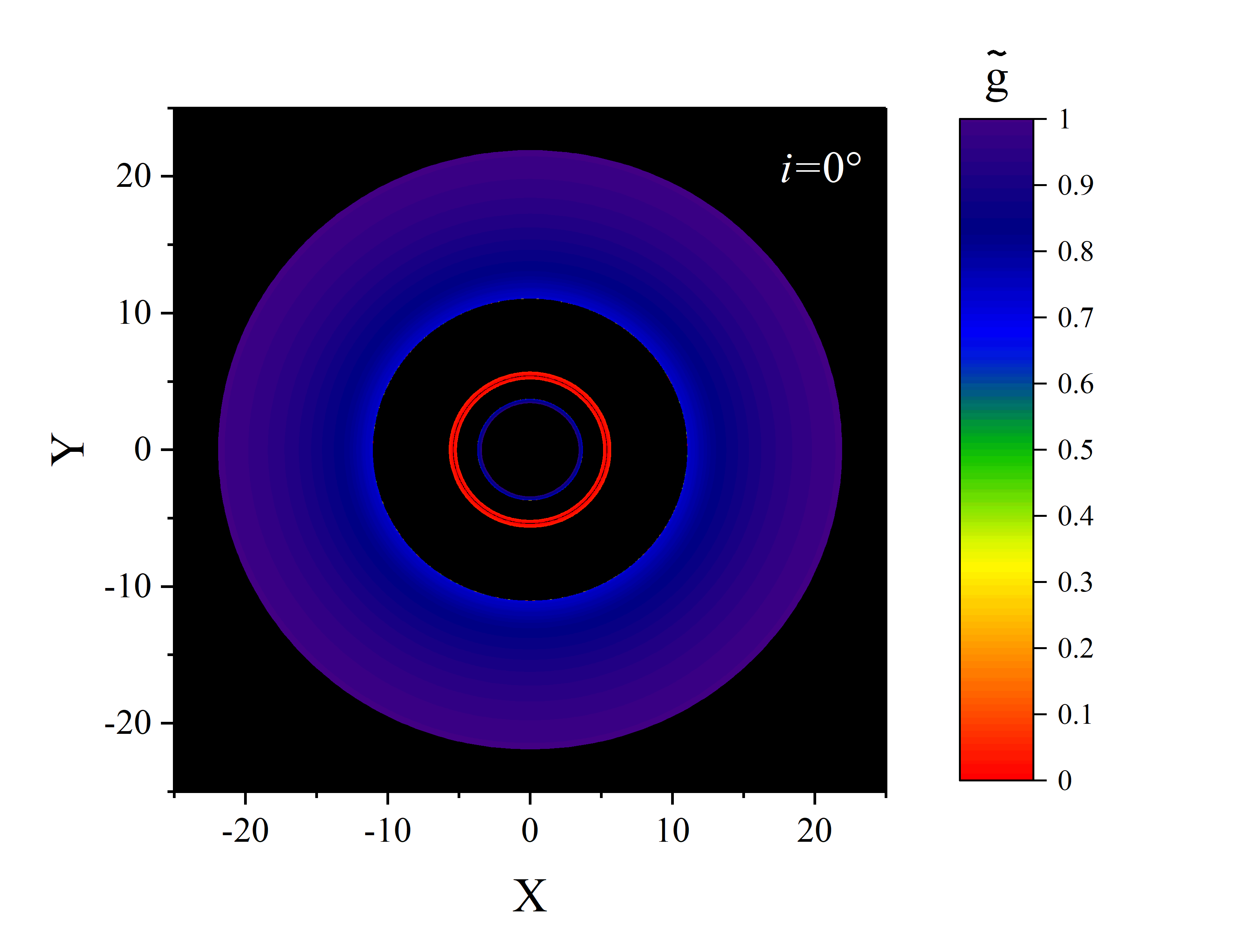}
\includegraphics[width=80mm]{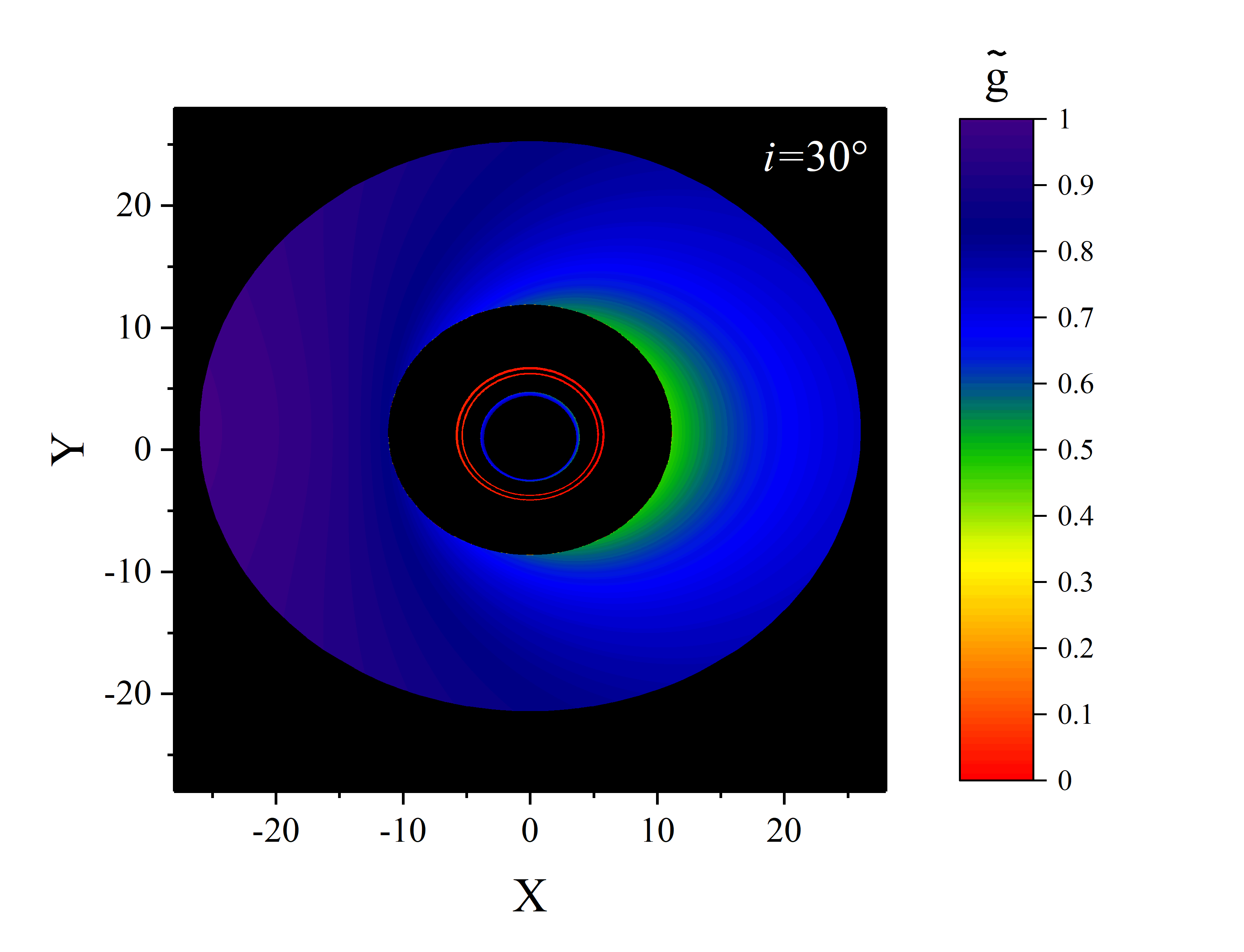}
    \caption{AD images for the  $S3$ type ($M=2$, $Q=0.99$, $n=14$, $\eta\approx 2.9998$) in full face and  for   inclination $i=30^o$.  The outer disk edge radius is $r=24$. SCO radii intervals are  $(0,0.22)$ (inner SCO  region), $(0.65,2.14)$ (second SCO ring), $(9,\infty)$ (outer unbounded SCO ring). Due to weak repulsion of the singularity  ($\eta\approx 2.9998$) the second  SCO rings is observed (red circles) and the inner SCO region is partially  invisible. }
    \label{fig:diskS3}
\end{figure}

\begin{figure}
    \centering
\includegraphics[width=80mm]{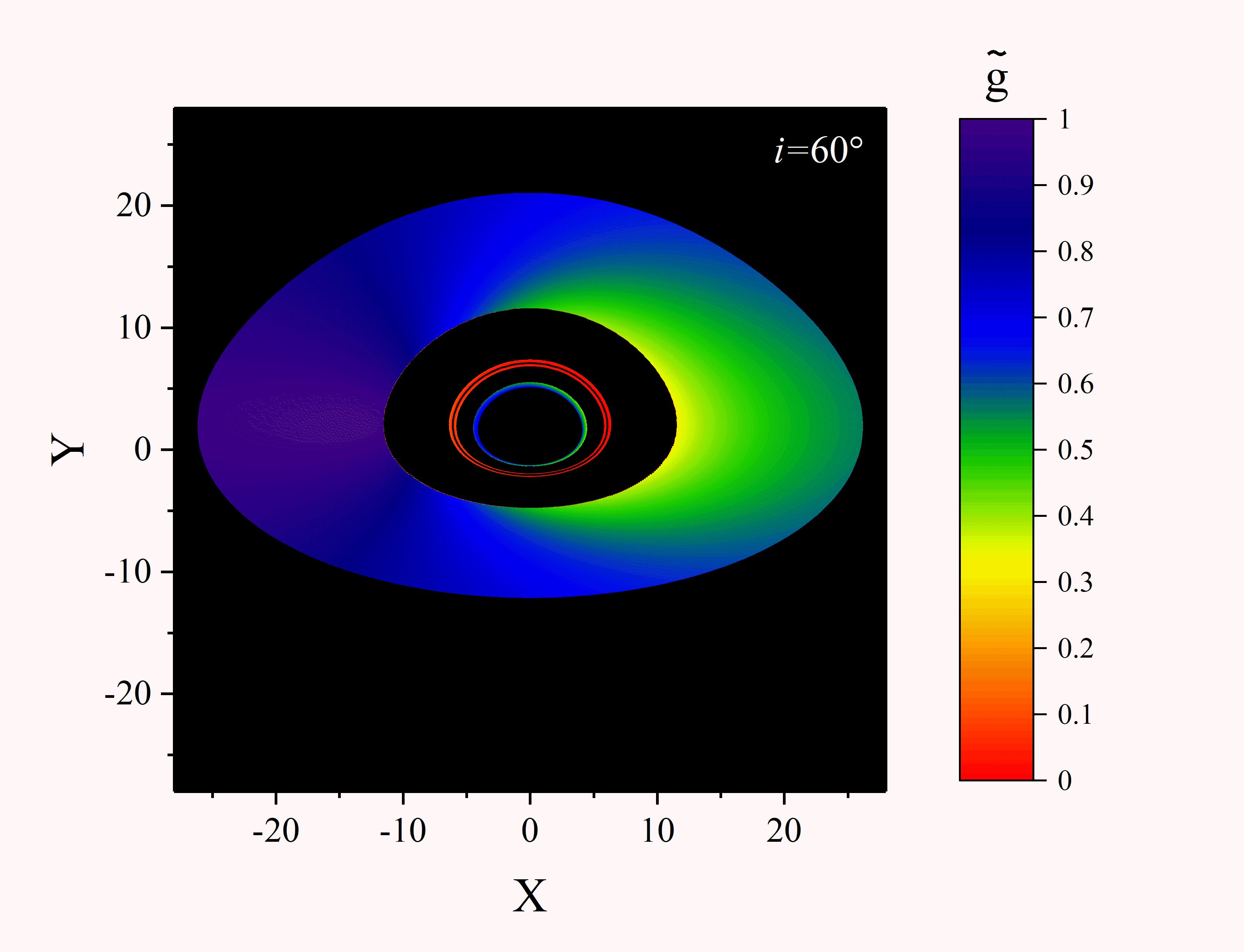}
\includegraphics[width=80mm]{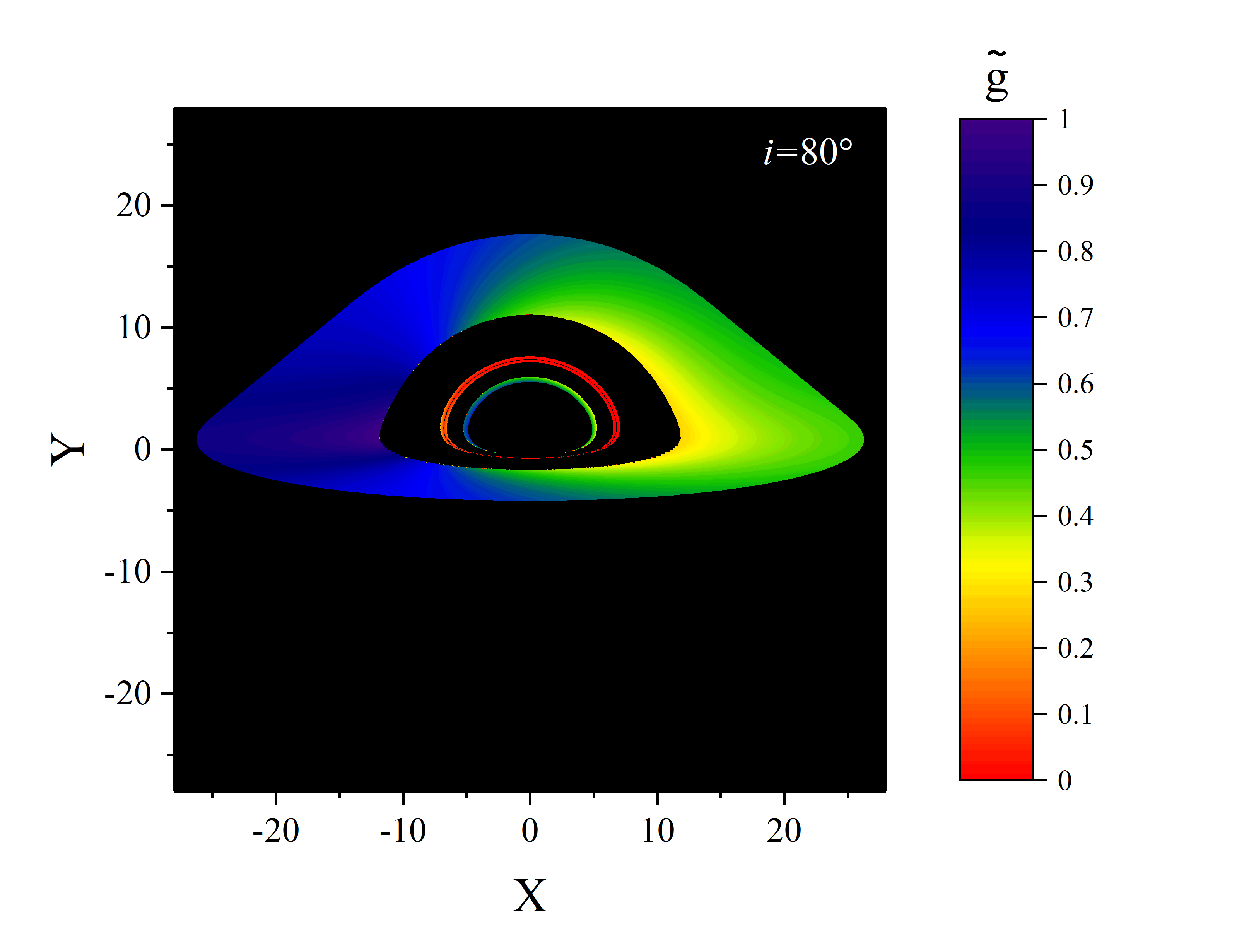}
    \caption{The same as on Fig. \ref{fig:diskS3} with inclinations $60^o$ and $80^o$.  }
    \label{fig:diskS3-b}
\end{figure}

Figs. \ref{fig:diskU1}--\ref{fig:diskS3-b} show the SCOD  contours  and the  distribution of $\tilde {g}$ over the image, visible from  infinity. The common feature of all the images is the existence of dark spot in the center like the ordinary black hole. This is either due to the properties of the photon trajectories falling from infinity ($\eta<3$, when these photons cannot reach the region near the center), or simply because of absence of SCO in this region.



\FloatBarrier
 \section{Discussion}\label{sec:discussion}
 We have studied isolated static spherically  symmetric configurations of General Relativity with minimally coupled nonlinear SF. The nonlinearity is introduced due to the SF potential $U(\phi)=\phi^{2n}$. For fixed $n>2$, we have shown that the solution of the corresponding  Einstein-SF system of equations exists and is unique under the  appropriate conditions  at spatial infinity describing an isolated object. This means that the configuration with scalar field is uniquely defined by two parameters: the configuration mass $M$ and the "scalar charge" $Q$ defined from the SF asymptotics $\phi(r)\approx Q/r$ for $r\to\infty$. 

  At the center we have a naked singularity with the asymptotic behaviour  that involves  parameter $\eta$, which describes the strength of the singularity \cite{ZhdSt}).
There is a critical value $\eta=3$ that separates two types of singularity -- attractive and repulsive -- with different behaviour of null geodesics near the center.  

The solutions of the Einstein-SF system have been investigated  numerically up to $n\sim 40$ for sufficiently large $M\sim 60$ and $Q\sim 60$. 
There are a lot of new  elements in comparison with FJNW, which arise in the dependencies of SCOD characteristics on  configuration parameters.  The most important difference from FJNW is associated with the emergence of the $S3$-type of SCOD  with  two rings of unstable circular orbits   for {$n\gtrsim 4.320$}. Indeed, though stable circular geodesics represent the simplified model of an accretion disk \cite{Page_Thorne}, the above stability properties can be important for real AD.

In  general, there are 4 possible SCOD types, the first three being similar to FJNW case. The $U1$ type is similar to the SCO distribution in the case of the Schwarzschild metric: there is an inner region around the center where circular orbits are either do not exist or are unstable, and there is an outer region of SCO that extends to infinity. For $S1$ type, SCO fill all the space starting from the center. For $S2$ type the stable orbits near the center are separated by a ring of unstable orbits from the outer SCO region that extends to infinity. Correspondingly,  the  images of the thin accretion disks are qualitatively similar to the FJNW case.  And, at last, there is a new type ($S3$) with an additional SCO region and with two rings of the unstable circular orbits that separate SCO rings.  Possible cases of SCOD are presented on Figs. \ref{fig:MQ1}, \ref{fig:MQ}  for different domains on the plane of the configuration parameters $M,Q$. One can infer that the  $S3$ type is less probable, moreover the innermost SCO rings typically have rather small  radii for moderate $M,Q$.

We plotted the  observable contours of the accretion disks of the same radius and images of SCO regions using the ray-tracing algorithm \cite{Psaltis_2011,JohannsenJ,Bambi_2012}. Figs. \ref{fig:diskU1}--\ref{fig:diskS3-b} also show the redshift distribution over AD image, which may be useful  to study deformation of the relativistic lines (e.g., Fe K$\alpha$) in the X-ray spectra of compact objects. 
A common feature of all the images is the dark spot in the center. This is either due to the absence of SCO near the center, or because strong bending of the photon trajectories near the naked  singularity. In case of $S2$ and $S3$ types, Figs. \ref{fig:diskS1}--\ref{fig:diskS3-b} demonstrate features around the center that are not observed for M87* shadow  \cite{Akiyama2019}. Apparently, these types should be ruled out in case of this object, though observations with better resolution are desirable to have a final answer.

To sum up, we note that basic  qualitative   properties of static spherically symmetric solutions of the Einstein-SF equations with  monomial potential (\ref{monomial_Self-int})  have much in common with the FJNW case \cite{Fisher, JNW}. However, there are subtle details in the distribution of matter around the configuration, which distinguish the  nonlinear SF.
\begin{acknowledgments}
O.S.S. and V.I.Z. acknowledge the support from National Research Foundation of Ukraine (project No. 2020.02/0073). The work of A.N.A. has been supported by a scientific program “Astronomy and space physics” of Taras Shevchenko National University of Kyiv (Project No. 19BF023-01).
\end{acknowledgments}
\bibliography{references.bib}
\appendix
\FloatBarrier
\section{Iteration method for solutions at large distances}\label{Iterations}
A consideration of existence and uniqueness for an isolated configuration in the asymptotically flat space time has been carried out in our  paper \cite{2019BTSNU..59....6A}, where we have used the so called "quasi-global" coordinates. Here we present a direct proof in the coordinate system defined by the metric representation (\ref{metric}).

	We introduce  variables
\[
X(r)= r \left(e^{-\beta}-1\right),\quad Y(r)= r^2 e^{\frac{\alpha-\beta}{2}}\frac{d\phi}{dr} 
\]
	The asymptotic flatness conditions (\ref{flattness},\ref{infinity_limit}) can be rewritten in terms of $X,Y,\alpha,\phi$ as
\begin{equation} \label{flattness-it}
\lim\limits_{r\to \infty}X(r)= -r_g,\quad
\lim\limits_{r\to \infty}Y(r)= -Q, \quad \lim\limits_{r\to \infty}\left[r\alpha(r)\right]= -r_g\,.
\end{equation}	
From (\ref{infinity_limit}) we also have
\begin{equation*} \label{infinity_limit-it}
\lim\limits_{r\to \infty} [r\phi(r)-Q]=
-\lim\limits_{r\to \infty} r \int\limits_{r}^{\infty}  \frac{x^2d\phi/dx+Q}{x^2}=0,
\end{equation*}	
then 
\begin{equation} \label{infinity_limit-it_phi}
\lim\limits_{r\to \infty} [r\phi(r)]=Q.
\end{equation}	
 Einstein equations (\ref{Ein_1-0},\ref{Ein_2-0}) and SF equation (\ref{equation-phi}) can be rewritten in terms of $Z\equiv \{X,Y,\alpha,\phi\}$ as the first order system. 
 
 Equations (\ref{Ein_1-0}, \ref{equation-phi}) yield
\begin{equation}
\label{Ein_1-0-iter}  
\frac{dX}{dr} =-8\pi \left[e^{-\alpha}\frac{Y^2}{2r^2}+r^2  |\phi|^{2n} \right] \,,
\end{equation}
\begin{equation} 
\label{equation-Dphi-iter}
\frac{dY}{dr}=2n r^2 e^{{\alpha}/{2}}\frac{ \phi|\phi|^{2n-2}}{\sqrt{1+X/r}}\, .
\end{equation}
In equation (\ref{Ein_2-0}) that takes the form  
\begin{equation*}
  \frac{d\alpha}{dr}= \frac{1}{1+X/r}\left\{-\frac{X(r)}{r^2}+8\pi   r \left[e^{-\alpha}\frac{Y^2}{2r^4}- |\phi|^{2n} \right]\right\} 
\end{equation*}
we separate out   the dominating term for $r\to\infty$:
\begin{equation}\label{Ein_2-0-iter}
  \frac{d\alpha}{dr}=
\frac{r_g}{r^2(1-r_g/r)}   +D(Z,r)\,,
\end{equation}
where
 \begin{equation} \label{D(Z,r)}
D(Z,r)= \frac{1}{1+X/r}\left\{-\frac{X+r_g}{r^2(1-r_g/r)}+8\pi   r \left[e^{-\alpha}\frac{Y^2}{2r^4}- |\phi|^{2n} \right]\right\}\,.
 \end{equation}
In the equation for $d\phi/dr$, which is expressed by means of  $Y$,   
\begin{equation*}  
  \frac{d\phi}{dr} =e^{ -\alpha/{2}}\frac{Y}{r^{2}\sqrt{1+X/r}}\,,
\end{equation*}
 we also separate out the dominating term:
 \begin{equation} \label{equation-phi-iter}
  \frac{d\phi}{dr}=-\frac{Q}{r^{2}\sqrt{1-r_g/r}}+ E(Z,r)\,,
\end{equation}
where
\begin{align*} 
&E(Z,r)=e^{ -\alpha/{2}}\frac{Y}{ r^2\sqrt{1+X/r}} +\frac{Q}{r^2 \sqrt{1-r_g/r}}=\\&=
\frac{Y\left(e^{ -\alpha/{2}}-1\right)}{r^2 \sqrt{1+X/r}}+  \frac{Q(X+r_g)}{r^3(\sqrt{1-r_g/r}+\sqrt{1+X/r})\sqrt{1-r_g/r}\sqrt{1+X/r} }+
\frac{Y+Q}{r^2\sqrt{1+X/r}}.
\end{align*}

Consider    set ${\bf S}$ of continuous vector-functions $Z(r)=\{X(r), Y(r),\alpha(r),\phi(r)\}$ satisfying 
\begin{equation}\label{Class_1}
|X(r)|\le 2r_g,\quad |\alpha(r)|\le 2r_g/r,\quad  |Y(r)|\le 2|Q|,\quad |\phi(r)|\le 2|Q|/r\,,\quad r\in [r_{\rm  in},\infty),
\end{equation}	
where $r_{\rm  in}>0$ will be further assumed to be sufficiently large. 
	
We shall construct a system of integral equations for solutions from ${\bf S}$, which is equivalent to equations (\cref{Ein_1-0-iter,Ein_2-0-iter,equation-Dphi-iter,equation-phi-iter}) with conditions (\ref{flattness-it}). For the  estimates below, it is essential that $n>2$; the case $n=2$ needs a separate consideration\footnote{This can be done in a similar way, but more cumbersome.} not presented here.	
 
Given the conditions  (\ref{flattness-it},\ref{infinity_limit-it_phi}), we get following system. 

Equations  (\ref{Ein_1-0-iter},\ref{equation-Dphi-iter}) yield
  \begin{equation}\label{Iter_1}
	X(r)=-r_g+A_1(Z,r),\quad A_1(Z,r)\equiv 8\pi\int\limits_{r}^{\infty}ds\, \left[\frac{Y^2(s)}{2s^2}  e^{-\alpha(s)}+s^2|\phi(s)|^{2n} \right],
	\end{equation}
	where we take into account $n>2$;
\begin{equation}\label{Iter_2}
	Y(r)=-Q+A_2(Z,r),\quad A_2(Z,r)\equiv -2n\int\limits_{r}^{\infty}ds\, \exp\left[\alpha(s)/{2}\right]\frac{s^2\phi(s)|\phi(s)|^{2n-2}}{\sqrt{1+X(s)/s}}\,.
	\end{equation}
Equations (\ref{Ein_2-0-iter},\ref{equation-phi-iter}) yield
\begin{equation}\label{Iter_3}
 \alpha(r)=\alpha_0(r)+A_3(Z,r)\,,\quad \alpha_0(r)=\ln(1-r_g/r)\,,\quad A_3(Z,r) \equiv -\int\limits_{r}^{\infty}D(Z,s) ds \, ;
 \end{equation}
\begin{equation}\label{Iter_4}
\phi(r)=\phi_0(r)+A_4(Z,r)\,,\quad \phi_0(r)=\frac{2Q}{r_g}\left[1-\sqrt{1-r_g/r}\right]\,,\quad A_4(Z,r)\equiv-\int\limits_{r}^{\infty}E(Z,s)ds \,.
\end{equation}
Here integral operators $A_i,i=1,...,4,$ are defined on ${\mathbf S}$.

Let  $Z(r)\equiv\{X(r), Y(r),\alpha(r),\phi(r)\}\in {\bf S}$. Denote
\begin{equation}\label{tilde_X_Y} 
\tilde 	X(r)=-r_g+A_1(Z,r),\quad \tilde	Y(r)=-Q+A_2(Z,r) .
\end{equation}

For  sufficiently large $r$,  simple estimates on account of (\ref{Class_1}) and $n>2$ yield
\begin{equation}\label{iter-esimates-A_1_A_3}
|\tilde X(r)+r_g|=
\left| A_1(Z,r)\right|\le \frac{C_1 }{r},\quad |\tilde Y(r)+Q|= \left| A_2(Z,r)\right|\le \frac{C_3 }{r^{2n-4}}, 
\end{equation}
Here and below $ C_i=C_i(M,Q),\,i=1,2,...$ are   finite positive constants. 
 For sufficiently large $r$ inequalities  (\ref{Class_1},\ref{iter-esimates-A_1_A_3}) yield
$|\tilde X(r)|\le 2r_g,\quad   |\tilde  Y(r)|\le 2|Q|$; whence 
$  Z'\equiv\{\tilde X(r),\alpha(r), \tilde Y(r),\phi(r)\}\in {\mathbf S}$.

  Now we denote
\begin{equation}	
\label{tilde_alpha}
\tilde  \alpha(r)=\alpha_0(r)+A_3(Z',r)\,, \quad \tilde \phi(r)=\phi_0(r)+A_4(Z',r). 
\end{equation}
Note that here we use $Z'$ instead of $Z$ and correspondingly   $\tilde X, \tilde Y$ from (\ref{tilde_X_Y}) instead of $\ X, Y$, which modifies the iteration scheme below.   This provides some technical convenience in view of the specific form of the equations  involved and allows us to avoid more stringent assumptions on ${\mathbf S}$.
Using (\ref{Class_1},\ref{iter-esimates-A_1_A_3}), for $n>2$ we have
\begin{equation}
    \label{iter-esimates-r_alpha+r_g}
|r\tilde \alpha(r)+r_g|=
\left| r A_3(Z',r)\right|\le \frac{C_4 }{r}.
\end{equation}
and 
\begin{equation} 
 |r\tilde \phi(r)+Q|= \left| r A_4(Z',r)\right|\le  C_5 \mu(r), 
\end{equation}
where $\mu(r)$ is defined immediately after formula (\ref{asymptotics_inf_metric1}).

Whence, for a sufficiently large $r$,  $|\tilde \alpha(r)|<2r_g/r$ and $|\phi(r)|<2|Q|/r$ that is    $\tilde Z\equiv\{\tilde X(r),\tilde Y(r),\tilde \alpha(r), \tilde \phi(r)\}\in{\bf S}$.
Therefore, we have mapping ${\mathbf R}:\,Z\to \tilde{Z}={\mathbf R(Z)}$ defined by (\ref{tilde_X_Y},\ref{tilde_alpha}), which transforms vector-function  $Z$  
\[Z\to Z' \to \tilde Z=\{\tilde X(r),\tilde Y(r),\tilde \alpha(r), \tilde \phi(r)\}\in{\bf S},\]
i.e. (for sufficiently large $r_{\rm  in}$),  operator ${\mathbf R}$  is correctly defined and  maps  functional class ${\mathbf S}$ into itself. Thus, initial equations    are  reduced to operator equation $
Z={\mathbf R}(Z)$.

Now we proceed to estimate contraction mapping properties of  ${\mathbf R}$. 
Let $Z_1=\{X_1,Y_1,\alpha_1,\phi_1\}\in {\mathbf S}$,  $Z_2= \{X_2,Y_2,\alpha_2,\phi_2\} \in {\mathbf S}$; $\tilde Z_1={\mathbf R}(Z_1),\quad  \tilde Z_2={\mathbf R}(Z_2)$;  $\delta Z\equiv Z_1-Z_2$. 

Denote
\begin{equation}\label{Class_2}
\|Z\|\equiv \sup\{|X(r)|+r|\alpha(r)|+|Y(r)|+r|\phi(r)|,\quad  r\in[r_{\rm  in},\infty).
\end{equation} 
Equations (\ref{tilde_X_Y}) on account of (\ref{Class_1}) yield
\begin{equation}\label{fp-1}
    |\delta \tilde X(r)|=|A_1(Z_1,r)-A_1(Z_2,r)|\le\frac{C_7}{r}\|\delta Z\|,
\end{equation}
\begin{equation}\label{fp-2}
 |\delta \tilde Y(r)|=|A_2(Z_1,r)-A_2(Z_2,r)|\le\frac{C_8}{r^{2n-4}}\|\delta Z\|.
\end{equation}
Here and below we systematically use the Lagrange finite-increments formula.
Using explicit form (\ref{D(Z,r)}) of $D(X,r)$   we have
\begin{align*} 
 &|D(Z_1,s)-D(Z_2,s)|\le\\& 
\le \frac{C_9}{s^2} |\delta  X(s)|+\frac{C_{10}}{s^3}  |\delta Y (s) |
 + \frac{C_{11}}{s^4} \left( s| \delta \alpha(s)| \right)+ \frac{C_{12}}{s^{2n-1}}\left( s|\delta  \phi(s)| \right).
\end{align*}
After substitution $Z_i\to\tilde Z_i,\, i=1,2$ on account of (\ref{fp-1},\ref{fp-2}) we have
\begin{equation*} 
|D(Z'_1,s)-D(Z'_2,s)|\le
\left\{ \frac{C_9 C_7}{s^3}   +\frac{C_{11}}{s^4} + \frac{C_{10} C_8 + C_{12}}{s^{2n-1}}\right\} \|\delta Z\| ,
\end{equation*}

whence
\begin{align}\label{fp-3}
|\delta \tilde \alpha(r)|=|A_3(Z'_1,r)-A_3(Z'_2,r)| \le \frac{C_{13}}{r^{2}}\|\delta Z\|\,.
\end{align}
Analogously
\begin{equation*} 
|E(Z_1,s)-E(Z_2,s)|\le
  \frac{C_{14}}{s^2}|\delta Y(s)|\left(1+O(1/s)\right)   
  +\frac{C_{15}}{s^4}|\delta X(s)| +\frac{C_{16}}{s^2}|\delta \alpha(s)|    ,
\end{equation*}
then
\begin{equation*} 
|E(Z'_1,s)-E(Z'_2,s)|\le
  \frac{C_{14}C_8}{s^{2n-2}}\|\delta Z\|\left(1+O(1/s)\right)   +\frac{C_{15} }{s^4}|\delta X(s)| +\frac{C_{16}}{s^3}|s\delta \alpha(s)| )  ,
\end{equation*}

\begin{equation}\label{fp-4}
  |\delta \tilde\phi(r)| =|A_4(Z'_1,r)-A_4(Z'_2,r)|\le \frac{C_{17}}{r}\mu(r)\|\delta Z\|.
\end{equation}
At last
\[\|\mathbf{R}(Z')-\mathbf{R}(Z)\|\le \max \left[\frac{C_{16}}{r_{\rm  in}},\frac{C_{17}}{r_{\rm  in}^{2n-4}}\right] \|\delta Z\|
\]
and we see that, for a sufficiently large $r_{\rm  in}$,    $\mathbf{R}$ is contraction mapping.
The solution can be obtained by successive approximations $Z_{(n+1)}={\mathbf R}(Z_{(n)})$ with zeroth iteration $Z_{(0)}=\{ -r_g,-Q,\alpha_0,\phi_0\}$.
Leaving the main terms for large $r$, from the first iteration we have
\begin{equation}\label{1st iter_X}
X_{(1)}(r)=-r_g+\frac{4\pi Q^2}{r}\left\{1+O[\mu(r)]\right\};
\end{equation}
\begin{equation}\label{1st iter_Y}
Y_{(1)}(r)=-Q-\frac{n Q |Q|^{2n-2}}{(n-2)r^{2n-4}}\left\{1+O[\mu(r)]\right\};
\end{equation}
\begin{equation}\label{1st iter_alpha}
\alpha_{(1)}(r)=\alpha_{{0}}+ O\left[ \frac{\mu(r)  }{r^2}\right]\,;
\end{equation}
\begin{equation}\label{1st iter_phi}
\phi_{(1)}(r)=\frac{Q}{r}\left\{1+\frac{r_g}{2r}+\frac{n|Q|^{2n-2}}{(n-2)(2n-3)r^{2n-4}}\right\}+ O\left[\frac{\mu(r)}{r^2}\right]\,.
\end{equation}
Here we have left those orders in $r^{-1}$ that will not change in subsequent iterations. These equations yield asymptotic relations (\ref{asymptotics_inf_metric},\ref{asymptotics_inf_metric1}).

\end{document}